\documentclass{aa}  

\usepackage{graphicx}
\usepackage{xcolor}
\usepackage{caption}
\usepackage[normalem]{ulem}
\usepackage{txfonts}
\usepackage{hyperref}
\hypersetup{
	colorlinks=true,
	urlcolor=blue,    % color of external links
	linkcolor=black,  % color of toc, list of figs etc.
	citecolor=blue,   % color of links to bibliography
}
\usepackage{pdflscape}   % recommended (rotates in PDF viewer)
\usepackage{eso-pic}

\newcommand{\LandscapePageNumber}{%
  \AddToShipoutPictureFG*{%
    \AtPageLowerLeft{%
      \raisebox{0.5\paperheight}{%
        \hspace{1.0cm}%
        \rotatebox{90}{\thepage}%
      }%
    }%
  }%
}

\usepackage{multirow}
\usepackage{array}
\usepackage{adjustbox}

\begin{document}

   \title{Two-body relaxation in the EMRI-TDE disk model for Quasi Periodic Eruptions}

   %\subtitle{}
   \titlerunning{Two-body relaxation and QPE}
   \authorrunning{Chiara Maria Allievi et al.}

   \author{Chiara Maria Allievi,
          \inst{1}
          Luca Broggi \inst{1}$^,$ \inst{2}
          Alberto Sesana, \inst{1}$^,$  \inst{2}
          \and
          Matteo Bonetti \inst{1}$^,$ \inst{2}$^,$ \inst{3}
          }

   \institute{Dipartimento di Fisica “G. Occhialini”, Università degli Studi di Milano-Bicocca, Piazza della Scienza 3, I-20126 Milano, Italy
         \and
             INFN, Sezione di Milano-Bicocca, Piazza della Scienza 3, I-20126 Milano, Italy
         \and
             INAF - Osservatorio Astronomico di Brera, via Brera 20, I-20121 Milano, Italy
             }

   \date{Received XXX; accepted YYY}

  \abstract
  {
   Quasi Periodic Eruptions (QPEs) are luminous bursts of soft X-rays recently discovered in galactic nuclei. They repeat on timescales of hours to weeks, superimposed to an otherwise stable quiescent X-ray level, consistent with emission from a radiatively efficient accretion flow around relatively low-mass massive black holes (MBHs). Although their physical origin is still debated, their quasi-periodicity naturally arises within the 'impact model', in which the X-ray bursts are generated by the interaction between a stellar black hole (sBH) or a star in a close orbit around the central MBH and the accretion disk formed by a tidal disruption event (TDE). 
   While this model is consistent with the phenomenology of QPEs, it remains unclear whether such specific physical configurations are sufficiently common to explain the observed QPE number density. 
   We present the first end-to-end quantitative calculation of the expected QPE rate and abundance within the framework of the impact model. To this purpose, we combine the rates of TDEs and extreme mass-ratio inspirals (EMRIs) around MBHs spanning a range of masses. We employ the public code \textsc{PhaseFlow} to simulate seven systems with MBH masses between $10^5 M_\odot$ and $10^8 M_\odot$, each surrounded by a three-component population: one composed of $1M_\odot$ stars, and two consisting of sBHs with masses of $10M_\odot$ and $40 M_\odot$. 
   Based on the emission constraints available in the literature, we restrict to sBH EMRIs on prograde orbit with eccentricity $e<0.5$ and inclination $i<20^{\circ}$ with respect to the accretion disk. For stellar EMRIs the constraints instead arise from the requirement that the star avoid tidal disruption. 
    We find that the predicted QPE number density spans the range $10^{-12} \rm Mpc^{-3}$ to $10^{-6} \rm Mpc^{-3}$, depending on the assumed orbital period interval and on the adopted eccentricity and inclination thresholds. QPEs generated by stellar EMRIs can reach number densities comparable to those inferred from current observations. In contrast, imposing the orbital constraints required for the sBH channel significantly suppresses the number of observable events, resulting in abundances approximately three orders of magnitude lower than in the stellar case. If the eccentricity and inclination constraints are relaxed, sBH-driven QPEs become compatible with the lower bound of the observationally inferred range. 
}

    \maketitle

\section{Introduction}

X-ray quasi-periodic eruptions (QPEs) are an extreme and rather enigmatic phenomenon identified in the last decade \citep{2019Miniutti}. These signals have luminosities up to $10^{43}$erg/s, a typical duration of one/few hours, and exhibit recurrence timescales around ten times longer. These eruptions are associated with fast spectral transitions between a cold and a warm phase in the accretion flow around a low-mass massive black hole (MBH). 

Following their discovery, QPEs have been observed so far from the nuclei of other eleven galaxies \citep{2020Giustini,2021Chakraborty,2021Arcodia,2023Quintin,2024Arcodia, Nicholl2024, 2024Guolo,2025Chakraborty,2025Hernandez-Garcia}.

The nature of the host galaxy is still under investigation. Half of them may be relatively low-mass
post-starburst galaxies, a population that might be a preferential host for observed tidal disruption events (TDEs), \citep{2019Miniutti,2021Chakraborty,2020Giustini,2021Arcodia,2024Guolo}. Despite very little X-ray obscuration, none of the QPE galaxies show the typical optical or UV broad emission lines associated with unobscured, type I active galactic nuclei (AGNs), although the observed narrow-lines suggest the presence of an ionizing AGN-like continuum \citep{2022Wevers}, possibly suggesting past AGN activity. 

The physical mechanism producing QPEs is the subject of active debate. A first class of models associates QPEs with intrinsic instabilities of the accretion flow around the central MBH. In this scenario, the eruptions arise from recurrent transitions in the inner disk, for instance due to thermal-viscous or radiation-pressure instabilities, which can drive repeated cycles of enhanced accretion and soft X-ray emission \citep{2022Pan,2023Sniegowska,2023Kaur}. These models naturally connect QPEs to the physics of accretion disks; however, because the instabilities are generated internally, they do not provide a robust external clock capable of enforcing the strict repeating periodicity observed in most QPE sources. As a result, they may struggle to reproduce features such as alternating long--short recurrence intervals and strong--weak amplitude patterns unless additional modulating physics is introduced.

A second class of models invokes the interaction between an accretion disk and a secondary object orbiting the MBH. In disk-impact models, a star or a stellar-mass black hole (sBH) on a bound orbit periodically crosses the disk. Each crossing shocks and strips part of the disk gas, producing expanding optically thick bubbles that cool and radiate in soft X-rays, giving rise to recurrent eruptions on a timescale related to the orbital period of the secondary \citep{2010Dai,2021Xian,2023Linial,2023Franchini}. In this framework, the disk may be supplied by a pre-existing weak AGN-like accretion flow or by the debris of a recent TDE. The latter possibility is particularly relevant because it links QPEs to transient disks formed after stellar disruptions \citep{2023Linial,2023Kaur}. Indeed, two known QPE sources show long-term decays in quiescent luminosity consistent with TDEs \citep{2023Miniutti,2024Arcodia}, and four observed TDEs have exhibited X-ray flares resembling individual QPE eruptions \citep{2021Chakraborty,2023Quintin,Nicholl2024,2025Chakraborty}. While TDEs are expected to occur in only about one out of every \(10^4\) galaxies, their detection in roughly half of the galaxies hosting QPEs suggests that such a connection might not be coincidental.

In this work we focus on the model for which the interaction between a star or a sBH and a TDE accretion disk gives rise to QPEs. Specifically, we build upon the results of \citet{2023Franchini}, who modeled successive passages of QPEs within the sBH EMRI–TDE disk framework, and we extend this model by also considering interactions involving stellar EMRIs. We present the derivation of the number density of QPEs in the Universe—referred to as the QPE volumetric abundance—to assess whether our theoretical predictions are consistent with observational estimates. 

The paper is organized as follows. In Sec \ref{sec:model} we review the aspects of the theory relevant to our model of the QPEs origin and we present the theoretical reasoning behind the computation of the cosmic QPE volumetric abundance. In Sec \ref{sec:simulations_of_two-body_relaxation} we present the simulations of the Nuclear Star Clusters (NSCs) we have taken into consideration and the implementation of the computation of the EMRI and TDE rates. In Sec \ref{sec:qpe_abundance} we explain the calculations behind the cosmic QPE volumetric abundance. Finally in Sec \ref{sec:results} we present the results of our work, so that we can discuss them in Sec \ref{sec:discussion} and draw our conclusions in Sec \ref{sec:conclusions}.

\section{Disk Impact Model} \label{sec:model}

\begin{figure}
    \centering
    \includegraphics[width=8cm]{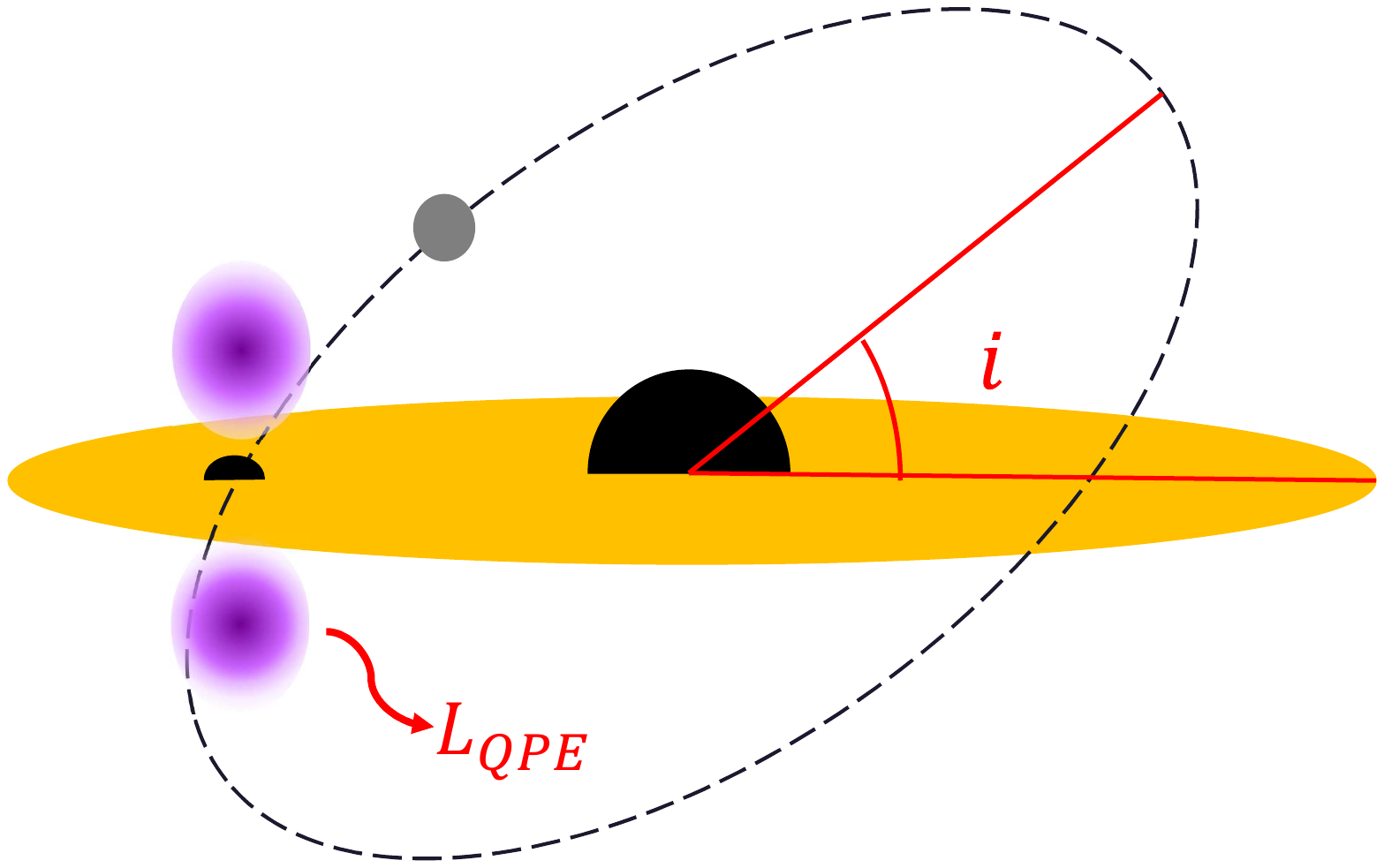}
\caption{ Schematic representation of the QPE disk-impact model. The central big black dot represents the MBH, the small gray dot represents the compact orbiter, either a sBH or a star, the yellow area represents the TDE accretion disk, and the dashed line represents the EMRI trajectory. The angle between the disk and the EMRI orbital plane is the inclination $i$. The violet bubbles represent the stripped mass that expands both above and below the disk after the impact. As soon as the bubbles become optically thin, they emit electromagnetic radiation ($L_{\mathrm{QPE}}$)}
    \label{fig:qpe_scheme}
\end{figure}

In the model we consider, QPEs are produced by the collisions between a TDE accretion disk and a compact orbiter\footnote{With “compact orbiters" we consider every object around the SMBH outside its tidal disruption radius or its loss cone.} (CO). Considering the scale of the system, the orbiter is typically undergoing an extreme mass-ratio inspiral (EMRI) around a MBH at the center of a galaxy (see Fig.\ref{fig:qpe_scheme}), as its motion is dominated by gravitational waves (GWs) emission. When the orbiter pierces the accretion disk, it strips some of its mass, which expands forming two bubbles: one on each side of the impact. Because of the expansion, the bubbles cool until they become optically thin and start emitting electromagnetic radiation. Since EMRIs can have semimajor axes entirely within the radial extent of the TDE disk, they periodically intersect the disk with a recurrence timescale that can match the observed timing of QPE bursts, including small differences among subsequent passages \citep{2023Franchini}. 

As sBHs do not have a proper physical surface, their cross section with the gas in the disk is set dynamically as the Bondi radius, i.e., $R = G M / (v_{\rm rel}^{2}+v_{\rm sound}^2)$ \citep{1952Bondi}. This implies some constraints on the EMRI orbit to reproduce the luminosities of QPEs, specifically the relative velocity between the sBH and the gas should be small for a sufficiently large radius. This translates into constraints on the orbital parameters, i.e., on the orbital eccentricity $e$ and the inclination $i$ of the orbital plane with respect to the disk: the eccentricity must be small to moderate ($e\lesssim0.5)$ and the inclination small ($i\lesssim20^\circ$) 
\citep{2023Franchini,2024Chakraborty}.

If the orbiter is a main sequence star, the impact radius is naturally set by the physical radius of the star. Therefore, the only constraint on the orbital properties arise from the stellar structure: the pericenter distance $r_p$ of the EMRI must be larger
than the tidal radius $r_t$, which encloses regions where the star is torn apart by the tidal field of the MBH. This requires $e<1-r_t/a$.

We estimate the abundance of QPE sources in the galactic population starting from the expected number of EMRI+TDE disk pairs in a single galactic nucleus. We assume that the compact orbiter is in place because of the combined effect of two-body relaxation and GW emission. In fact, weak scatterings can deflect the orbiters on low angular momentum orbits, where GWs emitted at periastron binds them to the central MBH initiating the slow GW driven adiabatic inspiral defining an EMRI.

The number of compact orbiters $N_\mathrm{CO}$ in the GW dominated region is estimated as the product between the rate at which orbiters enter this region of phase space and the coalescence time; this number must account for eccentricity and inclination constraints.
The number of TDE disks $N_\mathrm{disk}$ expected around the MBH is estimated as the product of the rate of stellar disruptions and the lifetime of the disk.
The product $N_\mathrm{CO}\times N_\mathrm{disk}$ gives the expected number of potential QPE sources in a given galactic nucleus. As this number is usually smaller than unity, we interpret it as the probability of observing an active QPE source when observing a galactic nucleus.

\subsection{TDEs and EMRIs from two-body relaxation}

Dynamical interactions within NSCs can drive stellar objects onto very tight orbits around the central MBH. Through this mechanism, some objects are driven either to plunge directly into the MBH, to enter an EMRI phase, or, if they are not compact enough to survive, to undergo a TDE. Specifically, the large, finite number of objects in NSCs induces small deviations from the smooth average potential of the system. As a consequence, the orbital parameters of a stellar object are subject to a Brownian motion known as two-body relaxation. The leading effect is the accumulation of distant two-body encounters between the considered stellar object and each of the others.
Considering the entire stellar population, the distribution of orbital parameters -- that in spherical symmetry can be identified with specific energy and specific angular momentum -- can be shown to evolve according to a diffusion-advection equation (the orbit averaged Fokker-Planck equation). Its coefficients are determined by the dynamics of two-body relaxation \citep{1987Spitzer}.

It is possible to modify the Fokker-Planck equation to account for disruptive processes like tidal disruptions or gravitational captures according to the loss cone theory\footnote{The name loss cone comes from the shape spanned by the velocity vectors that, at a given location in space, are tangent to orbits penetrating within a certain distance. The distance enclosing orbits leading to disruptions, or losses, is known as the loss cone radius.}. Loss cone phenomena are included by assuming an instantaneous disruption at the pericenter when it is smaller than a critical value $r_\mathrm{LC}$, which can be treated with modified boundary conditions for the equation \citep{1978Cohn}.

The Brownian motion of stars in angular momentum is generally faster than that in energy (especially for very eccentric orbits); an approximate description of relaxation is based on decoupling the two processes. One can use a simplified evolution equation for the distribution in energy (the ergodic Fokker-Planck equation), and assume at any instant the relaxed profile in angular momentum. The latter is generally approximated with the nearly-radial, steady-state profile induced by loss cone disruptions at fixed energy \citep{1979Cohn, 2013Merritt, 2017Vasiliev, 2020Stone}.

\subsubsection{TDEs and their disk}
A star can be tidally disrupted by the MBH if it enters a critical radius \citep{1975Hills,1988Rees,1989Phinney}. 
This distance is known as the tidal disruption radius $r_t$, and encloses the region in space where the tidal forces of the MBH overcome the self-gravity of a star, so that the latter is ripped apart. For a star of mass $M_\star$ and radius $R_\star$ it can be estimated as \citep{1975Hills}

\begin{equation}
    r_t = \eta_\star\,\left(\frac{M_\bullet}{M_\star}\right)^{1/3} R_\star 
    \label{eq:tidal_radius}
\end{equation}
where the form factor $\eta_\star\simeq1$ depends on the internal structure of the star and possibly the eccentricity of the orbit. For stars, we assume loss cone events to happen with $r_\mathrm{LC} = r_t$, with $\eta_\star=1$.

We assume that every star penetrating the tidal radius is disrupted by the MBH and forms an accretion disk. Simulations show that the stellar debris of a TDE follows a range of trajectories, with some of the material becoming unbound and escaping. Conversely, the bound debris return to the pericenter on highly eccentric orbits, possibly self-interacting because of relativistic precession and orbital dynamics. These collisions dissipate energy and facilitate the circularization of the debris, leading to the gradual formation of the accretion disk \citep{2010Dai, 2014Shen,2020Bonnerot,2021Curd}. This disk is the central engine powering the long term electromagnetic emission of a TDE, emitting radiation across multiple wavelengths as the material spirals inward and is accreted onto the black hole \citep{2010Lodato,2015Lodato,2020Mummery}. We assume a disk lifetime $\tau_\mathrm{disk}$ of 10 years, consistent with estimates from viscous disk-evolution models for TDE debris disks \citep{1990Cannizzo,2011Montesinos_Armijo,2024Piro}.

\subsubsection{EMRIs} \label{subsubsec:EMRI}

An EMRI can form when two-body relaxation brings a stellar object into an orbit so close to the central MBH, that the emission of GWs dominates over relaxation, driving the object to coalescence. 

The timescale for two-body relaxation $t_\mathrm{rlx}$ is set by the time it needs to change the angular momentum squared $J^2$ of an orbit by the order of itself. By defining eccentricity as $J^2/J_c^2 \equiv 1-e^2$, where $J_c(E)$ is the angular momentum of the circular orbit with energy $E$, one can write
\begin{equation}
    t_\mathrm{rlx} = \frac{1-e^2}{\mathcal{D}(E)} \,,
    \label{eq:t_rlx}
\end{equation}
where $\mathcal{D}$ is the diffusion coefficient determined by two-body relaxation and depends only on the orbital energy $E$. We opted for this definition of the eccentricity $e$ as it extends the Keplerian estimate to non-elliptical orbits influenced by the stellar potential.

\begin{figure}
    \centering
    \includegraphics[width=\linewidth]{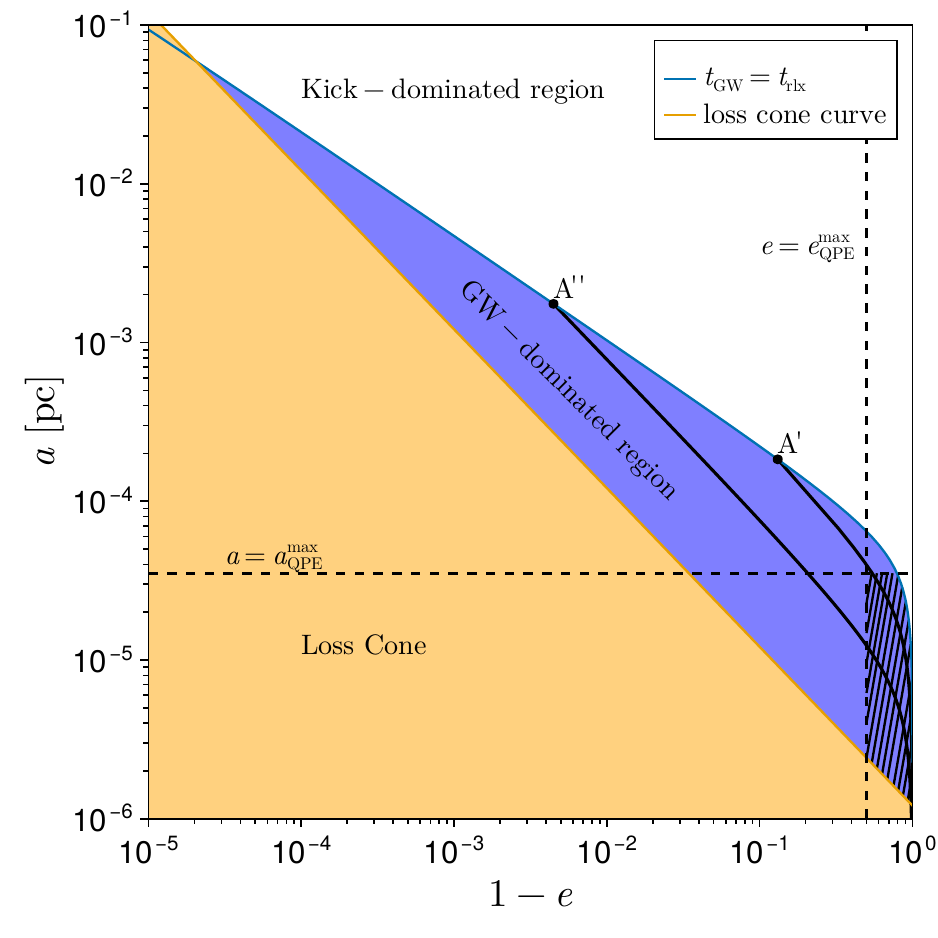}
    \caption{Trajectories of EMRIs dominated by GW emission. In the blue region the gravitational-wave timescale $t_\mathrm{GW}$ is shorter than the relaxation timescale $t_\mathrm{rlx}$; the orange area encloses plunging orbits; the white area encloses orbits dominated by two-body relaxation. 
    The shaded area encloses observable QPEs, that are determined by two parameters in our work: the maximum semi major axis $a_\mathrm{QPE}^\mathrm{max}$ (horizontal, dashed) and the maximum eccentricity $e_\mathrm{QPE}^\mathrm{max}$ (vertical, dashed).
    The black curves show the trajectory of sBHs that form EMRIs from Point A' and A'' and become observable QPEs.
    In this case, $M_\bullet = 3\times10^6M_\odot$, $M_\mathrm{CO} = 40M_\odot$. }
    \label{fig:separatrix}
\end{figure}

To estimate the timescale of GWs emission, we consider the smallest between the time required to change $1-e^2$ and $a$. The shortest is always $-a/\dot{a}_\mathrm{GW}$, that for the typically eccentric loss cone orbits can be estimated as \citep{1964Peters}:

\begin{equation}
    t_\mathrm{GW} = \frac{12\, \sqrt{2}}{85}\, \frac{c^5}{G^3\,M_\bullet^2\,M_\mathrm{CO}}\,a^4\,(1-e)^{7/2}
    \label{eq:t_gw}
\end{equation}
where we assumed that $M_\mathrm{CO} \ll M_\bullet$.
The dynamics of stellar objects can be described by delimiting two regions in the $(1-e,a)$ plane (see Fig.\ref{fig:separatrix}): \\
\textit{(i)} the region of stochastic evolution, where $t_\mathrm{rlx} < t_\mathrm{GW}$ and the orbital evolution is dominated by two-body encounters; \\
\textit{(ii)} the GW-dominated region, where $t_\mathrm{GW}< t_\mathrm{rlx}$, and the orbit evolution is dominated by the emission of GWs. Here the evolution is deterministic with increasingly good approximation as the CO's orbit shrinks.

Since relaxation for eccentric orbits changes $1-e^2$ faster than $a$, objects in the region dominated by two-body relaxation will be slowly pushed, approximately at fixed $a$, to pericenters $r_p<r_\mathrm{cap}$, resulting in gravitational captures. The value
\begin{equation}
r_\mathrm{cap}=8\,GM_\bullet/c^2 = 3.8\times 10^{-13}\, \mathrm{pc}\times(M_\bullet/M_\odot)
\label{eq:r_cap}
\end{equation}
gives, for very eccentric orbits, the value of the angular momentum of the geodesics entering the horizon of the MBH \citep{2013Merritt}. When including the emission of gravitational waves, it is possible to quantify the fraction of stellar objects entering $r_\mathrm{cap}$ directly (known as direct plunges), and those entering instead the GW-dominated region (thus forming an EMRI). With good approximation one can assume $r_\mathrm{LC}=r_\mathrm{cap}$, and count as direct plunges all the captures with semimajor axis larger than a critical value $a_c$, and count as EMRIs those with a smaller semimajor axis \citep{2005Hopman, 2016Bar-Or}. This picture fails for $M_\bullet \lesssim3 \times 10^5\,M_\odot$, as EMRIs can form with a large semimajor axis due to abrupt emission of GWs at periastron  \citep[known as cliffhangers, see][]{2024Qunbar, 2025Mancieri}. However, cliffanghers tend to be very eccentric and are unlikely to contribute significantly to the QPE population due to eccentricity constraints, as we will discuss later \citep{2025Mancieri2}. 
The critical semimajor axis can be estimated as the intersection between the line corresponding to $r_p = r_\mathrm{cap}$ and the line $t_\mathrm{rlx} = t_\mathrm{GW}$ (cf. Fig.~\ref{fig:separatrix}).
We will consider nuclei such that $r_t>r_\mathrm{cap}$, so that stellar EMRIs will be disrupted rather than being directly swallowed by the MBH.

\subsection{QPE abundance} \label{subsec:active_qpes}
We estimate the number of active QPEs per galaxy at any instant $t$ as $N_\mathrm{QPE}=N_\mathrm{disk}\, N_\mathrm{CO}$ \footnote{see Section~\ref{sec:model}}. To express $N_\mathrm{QPE}$ in terms of EMRI and TDE rates, we consider an observation starting at time $t$ and lasting $\Delta t_\mathrm{obs}$. During the observation time, we will see all the disks produced by disruptions that happened between $t-\tau_\mathrm{disk}$ and $t+\Delta t_\mathrm{obs}$. Similarly, the orbiter must be on an EMRI formed between $t-t_\mathrm{coal}$ and $t+\Delta t_\mathrm{obs}$ \footnote{We avoid introducing here the time required for light to reach the observer, as it is a pure time shift. Moreover, for this computation we assume that compatible EMRIs will form a QPE immediately after they form. In fact, over these timescales the EMRI rate changes negligibly.}, where $t_\mathrm{coal}$ is the time required for the sBH to coalesce or, when considering stellar EMRIs, the time required for a star to reach its tidal disruption radius. Since EMRIs are formed along the curve $t_\mathrm{rlx}=t_\mathrm{GW}$ and have different coalescence times, for each type of orbiter we introduce the average coalescence time $\bar{t}_\mathrm{coal} = N^{tot}_\mathrm{EMRI} / \Gamma_\mathrm{EMRI}.$

We assume that the EMRI rate $\Gamma_\mathrm{EMRI}$ and the TDE rate  $\Gamma_\mathrm{TDE}$ do not change significantly over the timescales we are considering, so that $\dot{\Gamma}_\mathrm{TDE}\,\tau_\mathrm{disk} \ll \Gamma_\mathrm{TDE}$ and $\dot{\Gamma}_\mathrm{EMRI}\,t_\mathrm{coal} \ll \Gamma_\mathrm{EMRI}$. We also introduce the QPE selection factor $\mu^{T,e,i}$, that is the fraction of compact orbiters with $T$, $e$ and $i$ compatible with detectable QPEs:
\begin{equation}\label{eq:mu}
    \mu^{T,e,i} = \frac{N_\mathrm{CO}}{N_\mathrm{EMRI}^{tot}} \, .
\end{equation}
The number of observed QPEs is:
\begin{equation}
\begin{split}
    N_\mathrm{QPE}^\mathrm{obs} &= N_\mathrm{disks}\,N_\mathrm{CO}\ \\  &= \int_{-\tau_\mathrm{disk}}^{\Delta t_\mathrm{obs}} \mathrm{d} t'\,\Gamma_\mathrm{TDE} \times \int_{-t_\mathrm{coal}}^{\Delta t_\mathrm{obs}} \mathrm{d} t'\,\Gamma_\mathrm{CO}\\
    &\simeq \Gamma_\mathrm{TDE} \, \Gamma_\mathrm{CO}\, (\Delta t_\mathrm{obs} + \tau_\mathrm{disk}) \, (\Delta t_\mathrm{obs} + t_\mathrm{coal}) \\
    &= \Gamma_\mathrm{TDE} \, \tau_\mathrm{disk}\, \Gamma_\mathrm{EMRI}\,\mu^{T,e,i}\,  \bar{t}_\mathrm{coal} \\
    &\quad + \Gamma_\mathrm{TDE} \, \Gamma_\mathrm{EMRI}\,\mu^{T,e,i} \, (\tau_\mathrm{disk} + \bar{t}_\mathrm{coal})\,\Delta t_\mathrm{obs} + \mathcal{O}\left(\Delta t_\mathrm{obs}^2\right)\,.
\end{split}
\end{equation}
We recognize the first term as the number of QPEs that were active at time $t$, and the second as the formation rate of QPEs times the observation time $\Delta t_\mathrm{obs}$.
Therefore, in this work we estimate the QPE abundance per galaxy as
\begin{equation}
    N_\mathrm{QPE} = \Gamma_\mathrm{TDE}\,\tau_\mathrm{disk}\,\Gamma_\mathrm{EMRI}\, \mu^{T,e,i} \, \bar{t}_\mathrm{coal} 
    \label{eq:active_qpe}
\end{equation}
and the QPE formation rate per galaxy as
\begin{equation}
    \Gamma_\mathrm{QPE} = \Gamma_\mathrm{TDE}\,\Gamma_\mathrm{EMRI}\, \mu^{T,e,i} \, (\tau_\mathrm{disk} + \bar{t}_\mathrm{coal})
    \label{eq:formation_rate} = N_\mathrm{QPE}\left(\frac{1}{\tau_\mathrm{disk}} + \frac{1}{\bar{t}_\mathrm{coal}}\right)\,.
\end{equation}

\begin{figure}
    \centering
    \includegraphics[width=\linewidth]{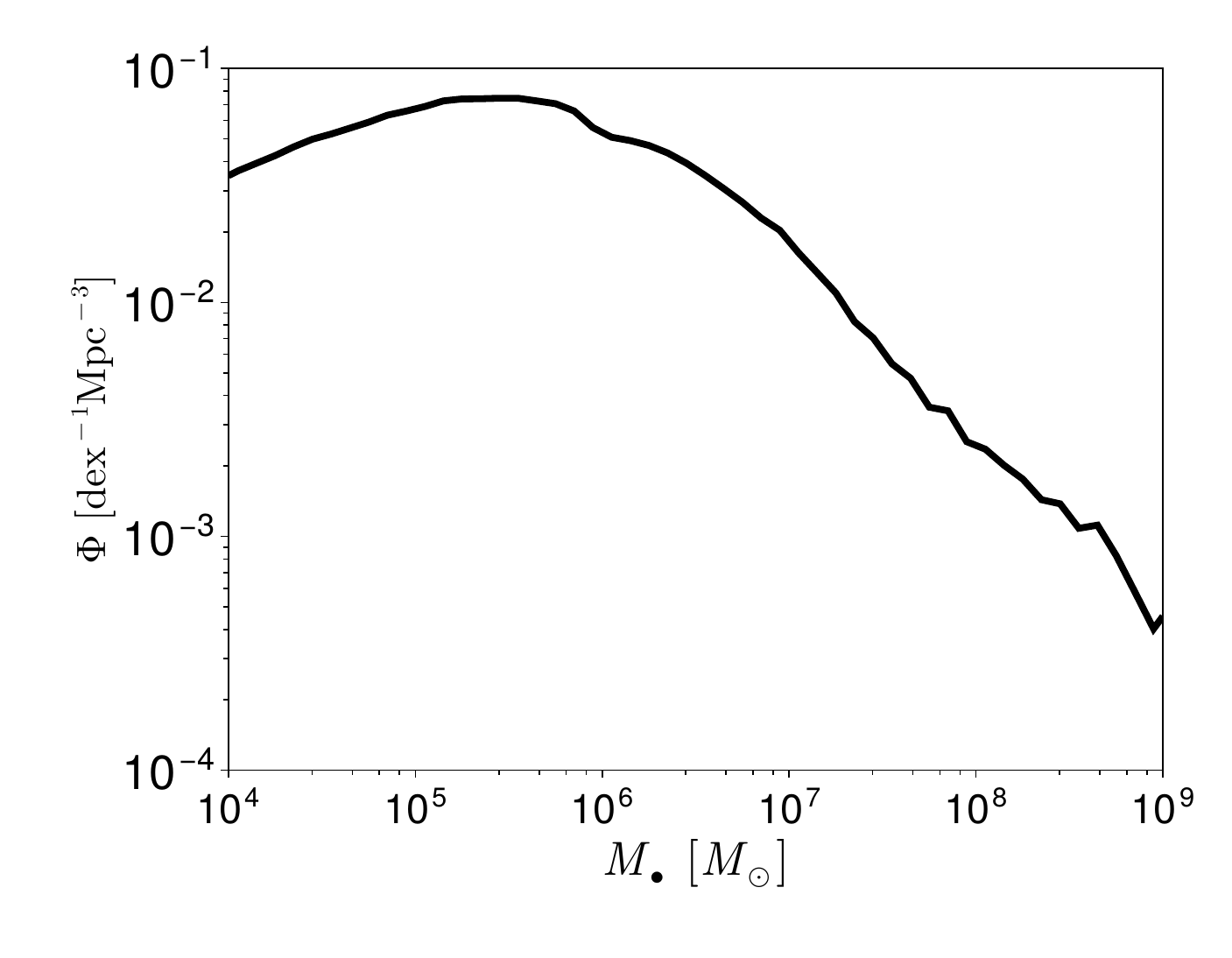}
    \caption{Black Hole Mass Function $\Phi$ used in this work \citep[from][]{2024Izquierdo-Villalba}.}
    \label{fig:bhmf}
\end{figure}

The cosmic volumetric abundance of QPEs is computed by multiplying the per-galaxy abundance, $N_\mathrm{QPE}(M_i, t)$ (Eq.~\eqref{eq:active_qpe}), by the local MBH mass function (that is, the number density of MBHs as a function of mass, see Fig.~\ref{fig:bhmf}) and integrating over $M_\bullet$. We caution that by doing so, we are implicitly assuming that TDE and EMRI rates are  determined solely by the MBH mass, and that all MBHs are surrounded by an NSC, which have been shown to dominate the production of TDEs \citep{2024Polkas}.

In practice, we simulate 0.5dex equally log-spaced values of $M_\bullet$ (cf. Sec.~\ref{sec:ic}), and we compute the number density of QPEs $n^i_\mathrm{QPE}$ expected from MBHs in the $i$-th mass bin as:

\begin{equation}
    n^i_{QPE} = N_\mathrm{QPE}(M_i, t) \; \int_{\Delta{M}_i} \! \mathrm{d} \log_{10}(M_\bullet) \; \Phi(M_\bullet),
    \label{eq:nqpe_bin}
\end{equation}

where $\Delta M_i$ is a 0.5dex bin centered in the i-th value of the simulated MBH mass.

Finally, to estimate the cosmic QPE volumetric abundance $\langle n_{QPE}\rangle$\, for each bin, we average $n^i_{QPE}(t)$ over the simulation and sum over all bins of $\log_{10} M_\bullet$ \footnote{$t + \Delta t_{\rm obs}=t_{\rm end}$}
\begin{equation}
    \left \langle n^i_\mathrm{QPE}\right \rangle = \frac{1}{t_\mathrm{end}} \int_0^{t_\mathrm{end}} \mathrm{d}t \, n^i_\mathrm{QPE}(t)
    \label{eq:nqpe_bin_avg}
\end{equation}
\begin{equation}
    \left \langle n_\mathrm{QPE}\right \rangle = \sum_i \left \langle n^i_\mathrm{QPE} \right \rangle \,.
    \label{eq:nqpe_total_avg}
\end{equation}

\section{Simulations of two-body relaxation and loss cone rates} \label{sec:simulations_of_two-body_relaxation}

For each system, we compute the rate of TDEs and EMRIs, and estimate the fraction of compact orbiters compatible with the disk-impact model for QPEs, by evolving the distribution function of stars and BHs in the Fokker-Planck formalism.

\subsection{\textsc{PhaseFlow} runs}

\textsc{PhaseFlow} is a publicly available code, that simulates the two-body relaxation of spherical isotropic stellar systems \citep{2017Vasiliev}. It solves the ergodic Fokker–Planck equation \citep{1980Cohn} and includes a sink term to account for loss cone captures. The one-dimensional approximation of relaxation in energy allows to span a wide range of parameters over very long evolution times, as the full 2D problem in energy and angular momentum is significantly more computationally expensive.

\begin{figure}
    \centering
    \includegraphics[width=9cm]{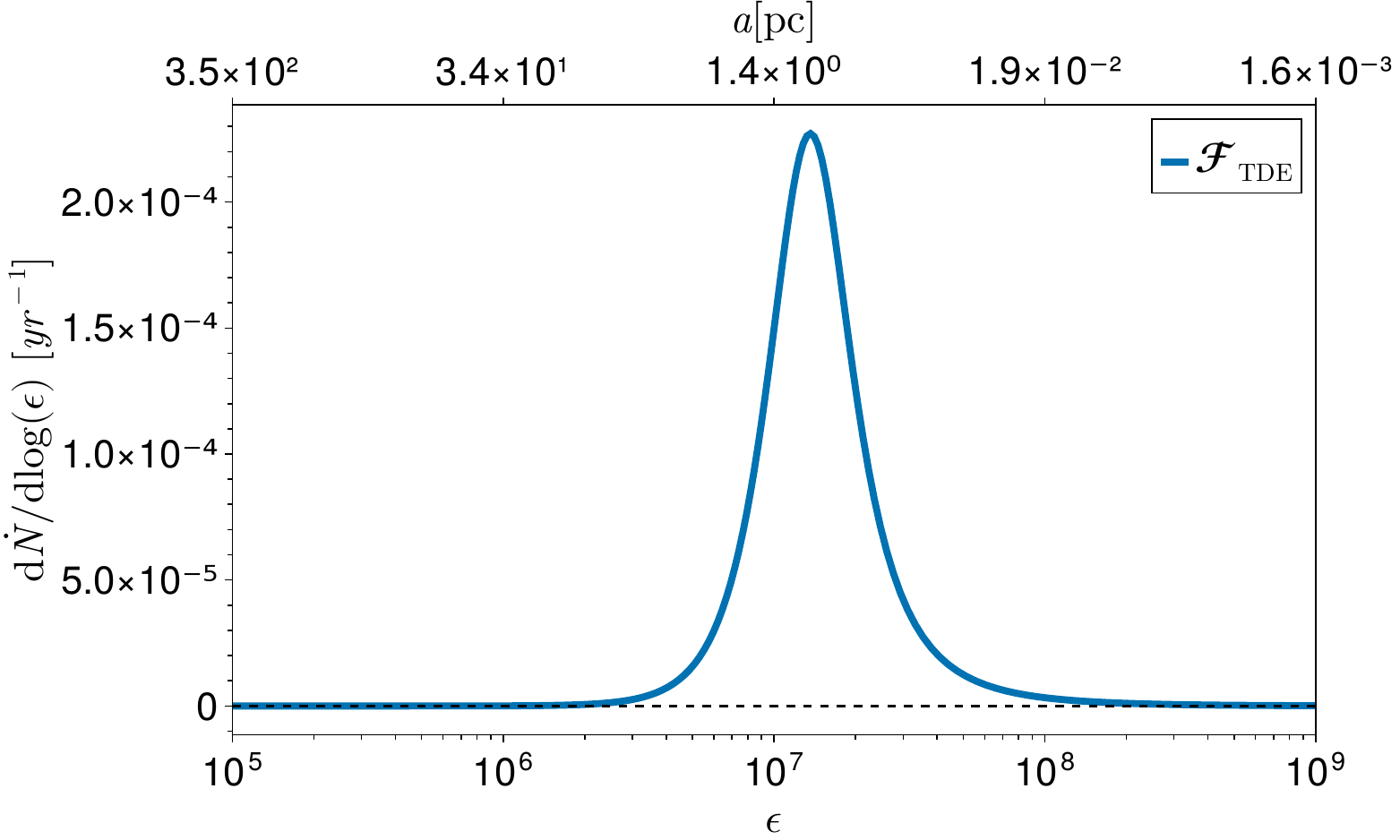}
    \caption{Differential rate of TDEs, shown through the related function $\mathrm{d}\dot N / \mathrm{d  log} \epsilon = \epsilon \mathcal{F}_\mathrm{TDE}$ as a function of the absolute value of the orbital energy $\epsilon=-E$ for a star with $M_\mathrm{CO} = 1M_\odot$, $M_\bullet = 3\times 10^6M_\odot$, at $t=8\times 10^6 yr$. The peak of $\mathrm{d}\dot N / \mathrm{d  log} \epsilon$ is located at $r_c(E) =$0.884 pc}
    \label{fig:tde_differential_rate}
\end{figure}

\subsubsection{Initial conditions}\label{sec:ic}

We consider systems with $\log_{10}M_\bullet/M_\odot = \{ 5,\ 5.5,\ 6,\ 6.5,\ 7,\ 7.5,\ 8\}$. For each MBH mass, we model the surrounding NSC with three stellar populations: one of stars and two of sBHs. The stellar component has total mass $20 M_\bullet$ and is composed of stars with mass $1M_\odot$. The sBHs components are formed by two populations of $10M_\odot$ and $40M_\odot$, respectively. Their abundance is chosen such that the $10M_\odot$ amount to a mass of $5/7\, M_\bullet$, while those with $40M_\odot$ represent a mass of $2/7 \, M_\bullet$\footnote{The proportions in total mass among the components match those obtained from the Kroupa Initial Mass Function \citep[see][]{2016Spera}}. As we anticipated, the capture radius of stars is, in the range of $M_\bullet$ we consider, the \textit{tidal radius} of Eq.~\eqref{eq:tidal_radius}. The capture radius for compact orbiters is the plunge radius Eq.~\eqref{eq:r_cap} (see Subsection \ref{subsubsec:EMRI}).

We set the size of the system so that the velocity dispersion $\sigma$ of stars, at the corresponding influence radius $r_h = G \, M_\bullet /\sigma^2$, is on the $M_\bullet-\sigma$ relation \citep{2000Ferrarese,2000Gebhardt}. We compute $\sigma$ according to the fit by \cite{2009Gultekin}:
\begin{equation}
    \sigma = 70\, \mathrm{km/s}\times \Bigg(\frac{M_\bullet}{1.53 \times 10^6 M_\odot} \Bigg)^{1 / 4.24}\, .
\end{equation}

We initialize each of the three components (stars, lighter sBHs, heavier sBHs) on  a Dehnen profile with inner slope 1.5
\citep{1993Dehnen, 1994Tremaine}:
\begin{equation}
    \rho = \rho_0 \, \left( \frac{r}{a} \right)^{-1.5} \left[ 1+  \left( \frac{r}{a} \right)\, \right]^{-2.5} 
\end{equation}
where $\rho_0$ is the density normalization; $a = 4\,r_h$ is the \textit{scale radius} ensuring the expected value of $\sigma$ at the influence radius $r_h$.

The relaxation timescale of the system depends on its properties: in our case the larger the total mass, the slower the relaxation. We set for the cases with $M_\bullet \leq 3\times10^6 M_\odot$ a maximum timestep of $10^6$ yr, and for the others a maximum timestep of $10^7$ yr. In all cases the total time of integration is $10^{10}$ yr, as a proxy of the age of the Universe.

\begin{figure}
    \centering
    \includegraphics[width=9cm]{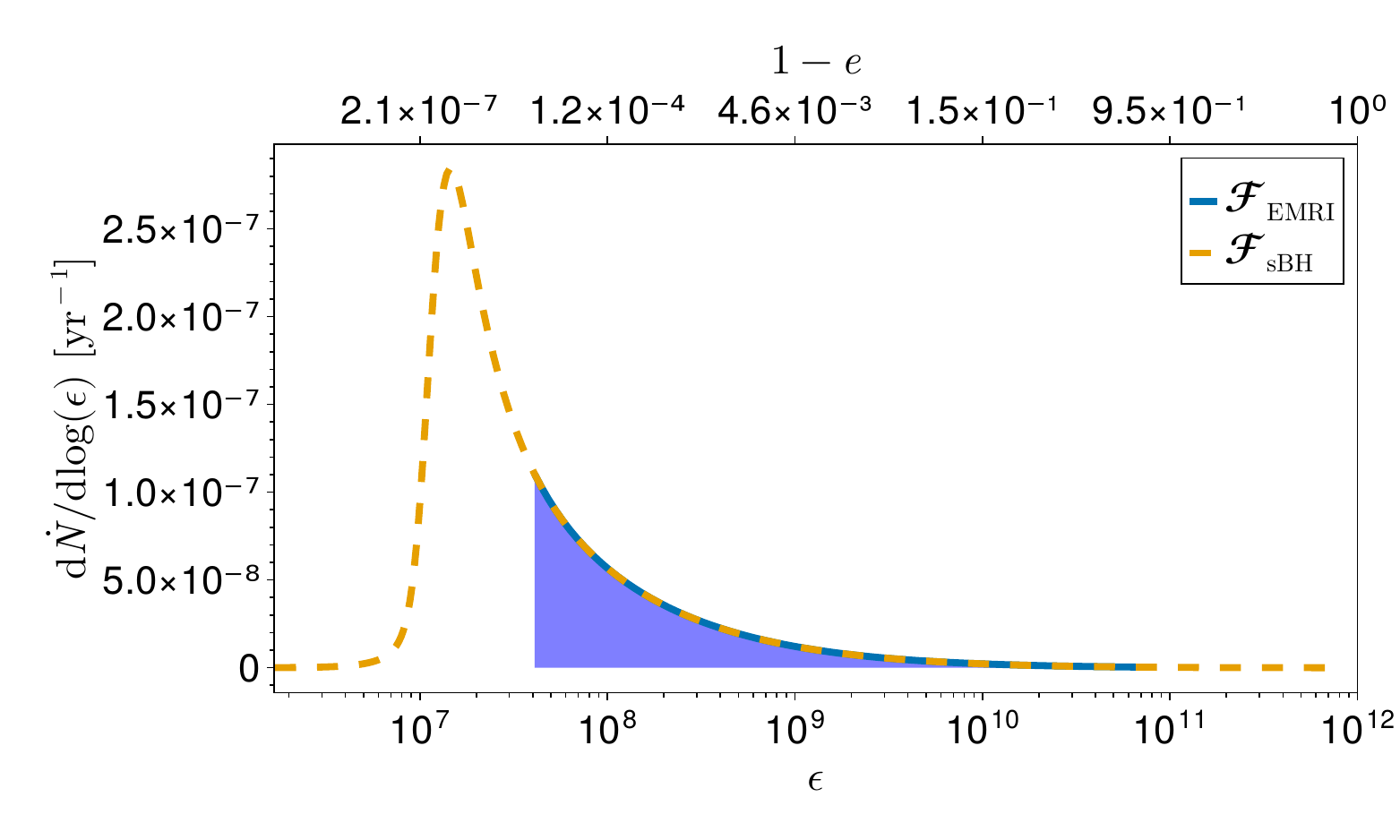}
    \caption{Differential rate of EMRIs $\mathcal{F}_\mathrm{EMRI}$ and direct plunges $\mathcal{F}_\mathrm{sBH}$ as functions of energy $\epsilon$ and eccentricity of the loss cone orbit $e$. We consider sBHs of $M_\mathrm{CO} = 40 M_\odot$ in a system with $M_\bullet = 3\times 10^6 M_\odot$ that has relaxed for $1$ Gyr. Our estimate of the EMRI rate is given by the light blue color-filled area. The eccentricity at the peak of the differential rate is $1-e=3.35\times10^{-7}$.
    }
    \label{fig:diff_rate_3x10^5}
\end{figure}

\subsection{TDE rates} \label{subsec:TDE_rates}

\textsc{PhaseFlow} tracks in time the number of stars entering the capture radius with a given energy $\mathcal{F}_\mathrm{TDE} = \mathrm{d}\,\Gamma_\mathrm{TDE} / \mathrm{d} E $ (see Fig.~\ref{fig:tde_differential_rate}). From this quantity, the TDE rates at time $t$ can be computed as:
\begin{equation}
    \Gamma_{TDE}(t) = \int_{-\infty}^{0} \mathrm{d} E \;\mathcal{F}_\mathrm{TDE}(t, E) .
    \label{eq:tde_rates}
\end{equation}
and therefore
\begin{equation}
    N_\mathrm{disks}(t) = \Gamma_\mathrm{TDE}(t) \, \tau_\mathrm{disk} \,.
    \label{eq:n_disk}
\end{equation}

\subsection{EMRI rates}\label{subsec:emri_rates}

In order to distinguish EMRIs from direct plunges, we compare the angular momentum relaxation timescale and the GW timescale of loss cone orbits\footnote{For a semi-analytical approach to the 2D Fokker-Planck equation, see \cite{2024Kaur}.}.  
Since it is used internally to compute the rate of disruptions, we recover the angular momentum relaxation time of loss cone orbits as computed by \textsc{PhaseFlow}. Specifically, \textsc{PhaseFlow} gives the diffusion coefficient $\mathcal{D}$, so that the relaxation time of an orbit is given by Eq.~\eqref{eq:t_rlx}.
The GW timescale is computed according to Eq.~\eqref{eq:t_gw}. Then, as presented in Subsection \ref{subsubsec:EMRI}, we select EMRIs as objects captured on orbits such that $t_\mathrm{GW}<t_\mathrm{rlx}$.

\begin{figure}
    \centering
    \includegraphics[width=9cm]{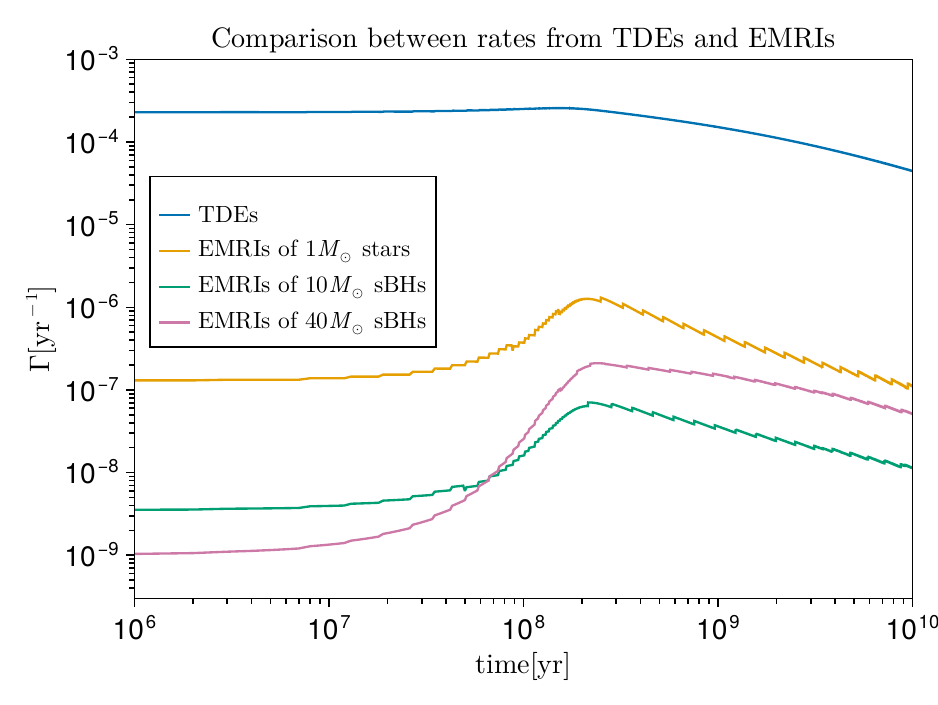}
    \caption{Time dependent EMRI rates for stars of $1M_\odot$ (orange), sBH of $10M_\odot$ (green) and $40M_\odot$ (pink), and TDE rates (blue), for $M_\bullet = 3\times10^6M_\odot$.}
    \label{fig:united_rates_316227.7660168379.png}
\end{figure}

Using the loss cone flux of a component $\mathcal{F}_\mathrm{CO} = \mathrm{d}\,\Gamma_\mathrm{CO} / \mathrm{d} E $ (see Fig.~\ref{fig:diff_rate_3x10^5}), we compute the rate of EMRI formation of that component as (see Fig.~\ref{fig:united_rates_316227.7660168379.png})
\begin{equation}
    \Gamma_\mathrm{EMRI}(t) = \int_{-\infty}^{E_\mathrm{GW}} \mathrm{d} E \;\mathcal{F}_\mathrm{CO}(t, E)
    \label{eq:tde_rates}
\end{equation}
where $E_\mathrm{GW}$ is the energy such that the curve $t_\mathrm{GW} = t_\mathrm{rlx}$ crosses the loss cone. The total number of EMRIs from that component must be computed by integrating over the $t_\mathrm{rlx}=t_\mathrm{GW}$ line as
\begin{equation}
    N^{tot}_\mathrm{EMRI} = \int \mathrm{d} E  \;\mathcal{F}_\mathrm{CO}(t, E) \, t_\mathrm{coal}(E)
\end{equation}
where $t_\mathrm{coal}(E)$ is the time-to-coalesce starting from the $t_\mathrm{rlx}=t_\mathrm{GW}$ curve with energy $E$.

\vspace{-5pt}

\section{Abundance and formation rate of QPEs} \label{sec:qpe_abundance}
\vspace{-0.2cm}

As we introduced in Sec.~\ref{subsec:active_qpes}, we compute the number of disks $N_\mathrm{disk}$ around the MBH by multiplying the instantaneous $\Gamma_\mathrm{TDE}(t)$ for $\tau_\mathrm{disk}$.
The number of compact orbiters, on the other hand, depends on the details of the model—and crucially on its constraints on inclination and eccentricity at the QPE semimajor axis. EMRIs form with typical semimajor axes in the range of $\sim10^{-5}-10^{-1}$pc, with periods much longer than those of observed QPEs. Assuming the Keplerian estimate\footnote{We assume that the $T_\mathrm{QPE}$ is half the orbital period, as for every radial period the sBH typically pierces through the accretion disk  twice.}
\begin{equation}
    a_\mathrm{QPE} \simeq \sqrt[3]{GM_\bullet\Bigg(\frac{T_\mathrm{QPE}}{\pi}\Bigg)^2}
    \label{eq:semi-major_axis}
\end{equation}
we consider for reference \footnote{We choose values corresponding to the recurrence period of AT2022upj (48 h) \citep{2025Chakraborty}, eRO-QPE3 (20 h) \citep{2024Arcodia}, GSN069 (9 h) \citep{2019Miniutti} and eRO-QPE2 (2 h) \citep{2021Arcodia}.}
\begin{align}
    a(48\mathrm{h}) &= 2.38 \times 10^{-7} \, \text{pc} \times \left( \frac{M_\bullet}{M_\odot} \right)^{1/3}\,,     \label{eq:AT2022upj} 
\\
    a(20\mathrm{h}) &= 1.34 \times 10^{-7} \, \text{pc} \times \left( \frac{M_\bullet}{M_\odot} \right)^{1/3}\,,     \label{eq:ero-qp3}
\\
    a(9\mathrm{h}) &= 7.5 \times 10^{-8} \, \text{pc} \times \left( \frac{M_\bullet}{M_\odot} \right)^{1/3} \,,    \label{eq:gsn069}
\\
    a(2\mathrm{h}) &= 2.9 \times 10^{-8} \, \text{pc} \times \left( \frac{M_\bullet}{M_\odot} \right)^{1/3}\,.
    \label{eq:ero-qp2}
\end{align}

\begin{figure}
    \centering
    \includegraphics[width=\linewidth]{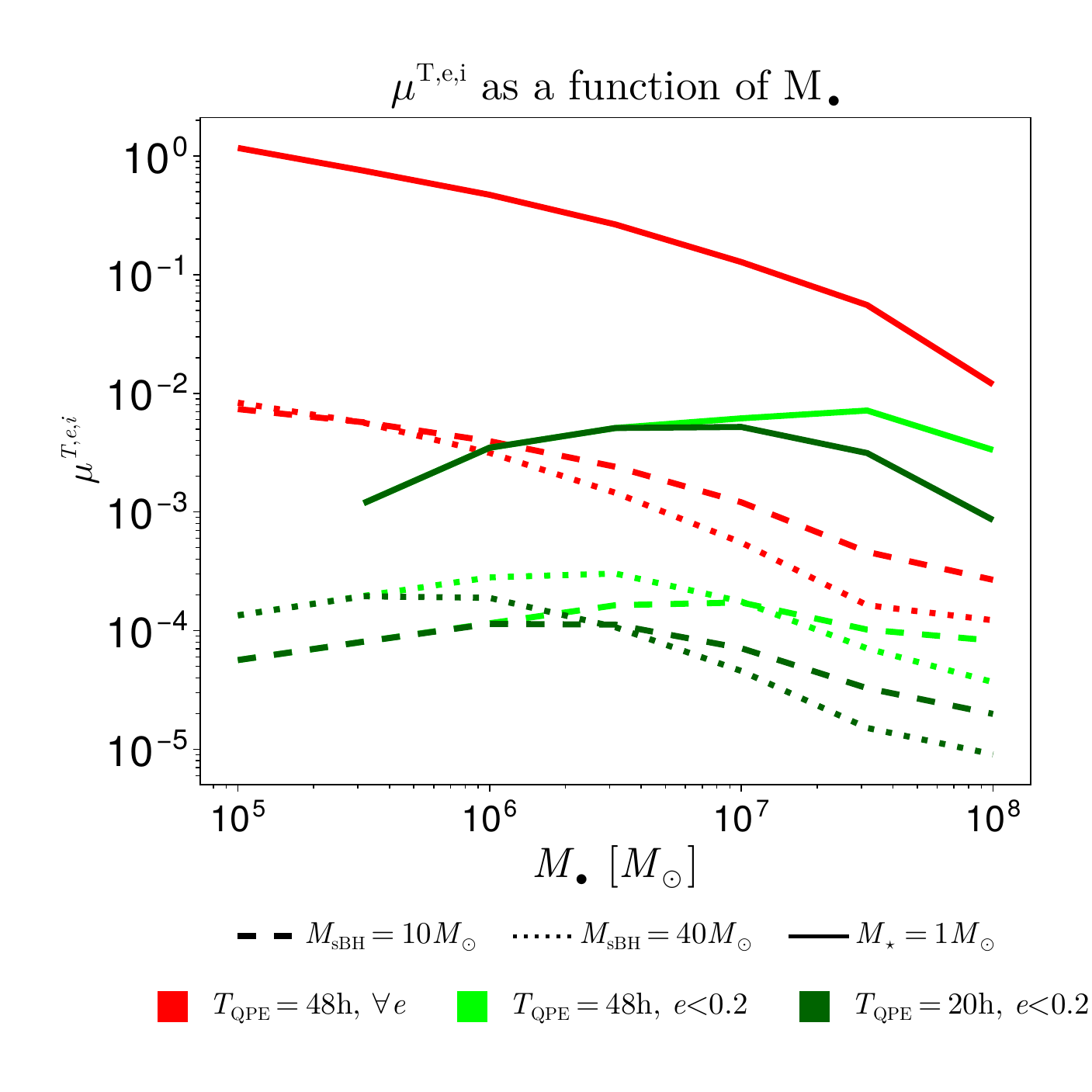}
    \caption{QPE selection factor $\mu^{T,e,i} = N_{CO} / N^{tot}_\mathrm{EMRI}$ as a function of $M_\bullet$, for different values of the maximum period $T_\mathrm{QPE}$, and the maximum eccentricity $e^\mathrm{max}_\mathrm{QPE}$ compatible with QPEs. }
    \label{fig:mu_MBH}
\end{figure}

\begin{figure*}
    \centering
    \includegraphics[width=0.48\linewidth]{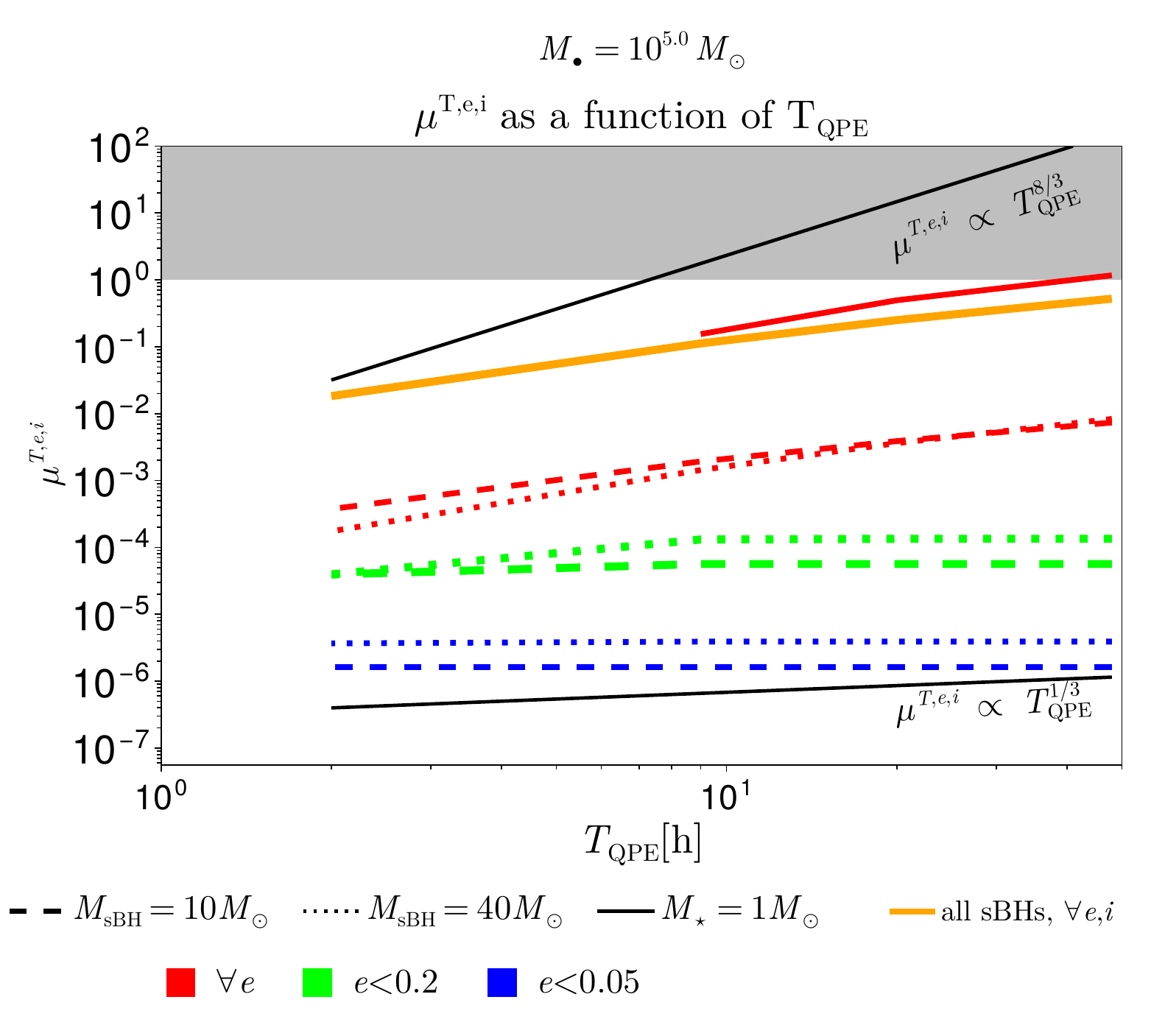} 
    \includegraphics[width=0.48\linewidth]{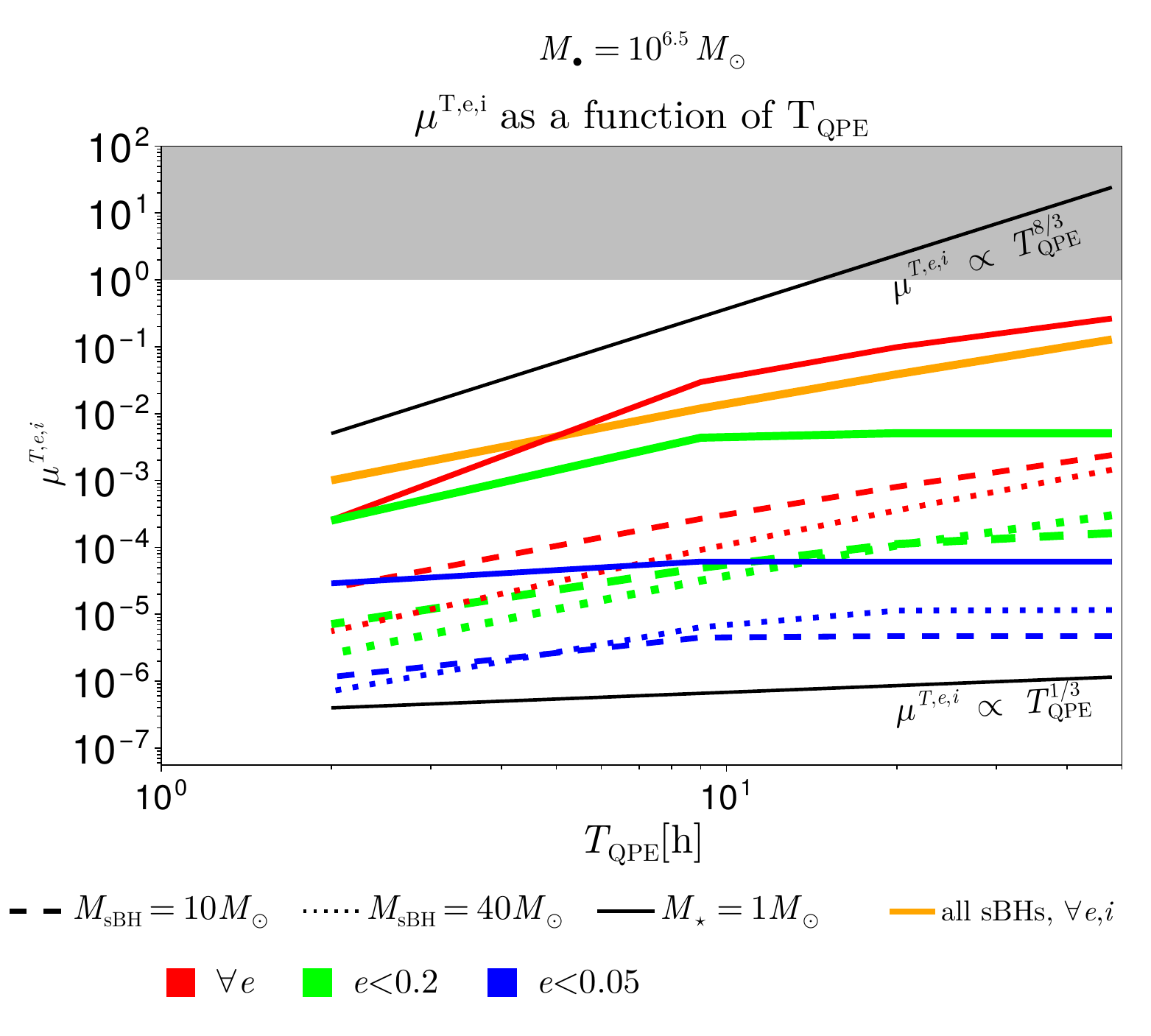}
    \caption{QPE selection factor $\mu^{T,e,i} = N_{CO} / N^{tot}_\mathrm{EMRI}$ for $M_\bullet=10^5M_\odot$ (left) and $M_\bullet=3\times 10^6M_\odot$ (right), for different values of the maximum period $T_\mathrm{QPE}$, and the maximum eccentricity $e^\mathrm{max}_\mathrm{QPE}$ compatible with QPEs. The gray region marks parameter values for which $\mu$ would exceed unity; these are unphysical because $\mu$ is defined as a fraction of all EMRIs (hence $0 \le \mu \le 1$). }
    \label{fig:mu_100000}
\end{figure*}

\subsection{Constraints on eccentricity and inclination} \label{subsec:e_constraints_EMRI}
Once we fix the maximum $a$ of observable QPEs $a^\mathrm{max}_\mathrm{QPE}$ and the maximum $e$ compatible with the model $e^\mathrm{max}_\mathrm{QPE}$, we evolve the EMRIs formed on the $t_\mathrm{rlx}=t_\mathrm{GW}$ line up until their $a$ and $e$ are both within the constraints (cf. black dashed area in Fig.~\ref{fig:separatrix}). While $a^\mathrm{max}_\mathrm{QPE}$ can be constrained observationally from the typical recurrence timescales (through the period--semimajor-axis mapping), the same is not true for $e^\mathrm{max}_\mathrm{QPE}$. 
The relevant eccentricity threshold depends on the (still uncertain) physical mechanism responsible for QPE emission, and therefore we treat $e^\mathrm{max}_\mathrm{QPE}$ as a model parameter.

We use the following strategy to compute the number of active QPEs from a snapshot of \textsc{PhaseFlow}. For each energy bin along the $t_\mathrm{rlx}=t_\mathrm{GW}$ line, we consider its center as the initial point of an EMRI trajectory (see Fig.~\ref{fig:separatrix} for a visual reference). From the 2.5 post-Newtonian orbit-averaged trajectory \citep{1964Peters} we find the eccentricity $e_1$ such that $a(e_1) = a^\mathrm{max}_\mathrm{QPE}$, where 

\begin{equation}
\label{eq:final_e}
    a(e) = c_0 \, \frac{e^{12/19}}{1-e^2} \left[1+\frac{121}{304} e^2\right]^{870/2299} \,.
\end{equation}
where $c_0$ is determined by the initial condition of $a$ and $e$.
If $e_1<e^\mathrm{max}_\mathrm{QPE}$, the trajectory is compatible with QPE orbiters from $(e_1, a^\mathrm{max}_\mathrm{QPE})$ to coalescence. If $e_1 > e^\mathrm{max}_\mathrm{QPE}$, we further evolve the trajectory up to $a_1 = a( e^\mathrm{max}_\mathrm{QPE} )$; this trajectory is compatible with QPE orbiters from $(e^\mathrm{max}_\mathrm{QPE}, a_1)$ until coalescence.

We show detailed examples of the possible evolution in phase space of ($1-e,a$) for two sBHs, as well as the selected $(1-e, a)$ parameters of QPEs given $(1-e^\mathrm{max}_\mathrm{QPE}, a^\mathrm{max}_\mathrm{QPE})$, in Fig.~\ref{fig:separatrix}.

The process of two-body scattering is treated under the assumption of spherical symmetry. 
In a spherically symmetric potential around a non-spinning MBH, we expect the orbital angular-momentum directions of bound sBHs and stars to be isotropically distributed, i.e., with random orbital inclination $i$. This assumption may break down if MBH spin (Kerr geometry) and relativistic precession introduce a preference for prograde or low-inclination orbits, potentially modifying EMRI rates and the inclination distribution. Accordingly, the direction of angular momentum is set to be uniformly distributed on the sphere at the beginning of the GWs dominated phase.
The probability that the relative $i$ between the TDE disk and the EMRI orbit is smaller than $i^\mathrm{max}_\mathrm{QPE}$ is: 
\begin{equation}
    P(0<i<i^\mathrm{max}_\mathrm{QPE}) = \frac{1}{2} \int^{i^\mathrm{max}_\mathrm{QPE}}_{0} d\theta \sin{\theta}
    \label{eq:inclination}
\end{equation}
and for sBHs we choose a typical value $i^\mathrm{max}_\mathrm{QPE} = 20^\circ$, giving $P(i<i^\mathrm{max}_\mathrm{QPE})=0.03$. We set no constraints on the inclination of stellar EMRIs. 

\subsection{Coalescence time and $N_\mathrm{CO}$} \label{subsec:t_coal}

Considering that EMRI formed with energy $E$ are compatible with QPEs only from $(e_1, a^\mathrm{max}_\mathrm{QPE})$ or $(e^\mathrm{max}_\mathrm{QPE},a_1)$, to compute $N_\mathrm{CO}$ one must compute the coalescence time from that location. We estimate it as \citep{1964Peters}
\begin{equation}
    t^\mathrm{QPE}_{coal}(E) = \frac{12}{19}\frac{c_0^4}{\beta}\times\int^{e_\mathrm{QPE}}_{e_\mathrm{fin}}\frac{de \ e^{29/19}[1+(121/304)e^2]^{1181/2299}}{(1-e^2)^{3/2}}
    \label{eq:t_coal}
\end{equation}
where $c_0$ is the quantity appearing in Eq.~\eqref{eq:final_e}, while $e_\mathrm{QPE}=e_1$ or $e^\mathrm{max}_\mathrm{QPE}$, and $e_\mathrm{fin}$ is the eccentricity at merger or disruption. Consequently, $r_\mathrm{cap}=a(e_\mathrm{fin})(1-e_\mathrm{fin})$ for sBHs, while for stars is the solution of $a(e_\mathrm{fin}) (1-e_\mathrm{fin}) = r_t$ using Eq.~\eqref{eq:final_e}.

We can finally assemble the ingredients to compute the expected number of orbiters compatible with the QPE model
\begin{equation}
    N_\mathrm{CO}(t) = \int \mathrm{d}E \, \mathcal{F}_\mathrm{EMRI}(t,E) \, t_\mathrm{coal}^\mathrm{QPE}(E) \,.
    \label{eq:n_co}
\end{equation}
Then, using Eqs.~\eqref{eq:n_co} and \eqref{eq:n_disk}, we compute the number of active QPEs as $N_\mathrm{QPE} = N_\mathrm{disks}\, N_\mathrm{CO}$.

\section{Results} \label{sec:results}

We summarize in this section the results of our analysis on the number of QPEs in the NSCs we consider, and on the inferred cosmic abundance of QPEs.

In Table~\ref{table:1} of Appendix~\ref{app:tables} we present the average QPE selection factor $\mu^{T,e,i}$, which is the fraction of EMRIs compatible with QPEs over $10$ Gyrs of simulation. We remark that it is defined in terms of three parameters: the maximum inclination $i^\mathrm{max}_\mathrm{QPE}$, the maximum eccentricity $e^\mathrm{max}_\mathrm{QPE}$, and the maximum orbital period $T_\mathrm{QPE}$. Even if the disk-impact model does not impose direct constraints on the eccentricity of stellar orbiters, the impact of $e^\mathrm{max}_\mathrm{QPE}$ is non-trivial, so we include it for the sake of completeness. We set $i^\mathrm{max}_\mathrm{QPE}=20^\circ$ for sBHs and no constraints on $i$ for stars. The scalings of $\dot{a}$ and $\dot{e}$ impact directly the trend of $\mu^{T,e,i}$ because of GW emission. At fixed constraints on $e$ and $T$, the fraction of COs compatible with QPEs decreases when $M_\bullet$ increases for large $M_\bullet$, as $T_\mathrm{QPE}$ restricts to a smaller and smaller fraction of EMRIs. At small $M_\bullet$ the circularization due to GW emission can reduce $\mu^{T,e,i}$, especially in the case of sBHs (see Fig.~\ref{fig:mu_MBH}). The number of objects with semimajor axis smaller than $a$ behaves like $t_\mathrm{GW}$ \citep[e.g.,][]{2017Amaro-Seoane}, setting the scaling of $\mu^{T,e,i}$ for $e^\mathrm{max}_\mathrm{QPE}=1$. As $t_\mathrm{GW}\sim a^{1/2}\sim T^{1/3}$ for very eccentric orbits and $t_\mathrm{GW}\sim a^4\sim T^{8/3}$ for nearly circular orbits, the scalings we find are bound by these two extremes. Due to the mixed eccentricity distribution, $\mu^{T,e,i}$ is almost linear with $T_\mathrm{QPE}$ at $M_\bullet=10^5\,M_\odot$, while for larger $M_\bullet$ it becomes steeper (see Fig.~\ref{fig:mu_100000}). In general, most EMRIs are found at moderate eccentricities, so that the impact of the eccentricity cut is evident for both sBHs and stars: larger $e_\mathrm{max}^\mathrm{QPE}$ gives larger $\mu^{T,e,i}$, although the magnitude of the change depends on the mass of the stellar object and the NSC. $\mu^{T,e,i}$ is generally comparable among the two species of sBHs, but is typically larger for stars because there is no inclination constraint and they are lighter. In fact, the GW dominated region for stars is smaller, and a larger fraction corresponds to periods compatible with QPEs. However, for large $M_\bullet$ the period at $r_t$ is so long that  $\mu^{T,e,i}=0$ for the shortest values of $T_\mathrm{QPE}$. For SMBH masses of $3\times 10^7\,M_\odot$ and $10^8\,M_\odot$, the same happens when the period at $r_\mathrm{cap}$ is longer than $T_\mathrm{QPE}$ (see Table~\ref{table:1} of Appendix~\ref{app:tables}).

In Figs.~\ref{fig:QPE_rates_1} and \ref{fig:QPE_rates_2} we show the number of active QPEs $N_\mathrm{QPE}$ for two representative systems in time. Since it can be expressed in terms of the abundance (Eq.~\eqref{eq:formation_rate}), assuming $\tau_\mathrm{disk} \ll \bar t_\mathrm{coal}$, the formation rate is
$\Gamma_{\rm QPE} =N_{\rm QPE} /{\tau_\mathrm{disk}}$, and the two quantities follow the same trend.
At $t\gtrsim 1$ Gyr, $N_\mathrm{QPE}$ starts to decline, following the long term evolution of loss cone event rates.
We report in Table~\ref{table:2} of Appendix~\ref{app:tables} the time average value $\langle N_\mathrm{QPE} \rangle$ over $10$ Gyr for all the NSCs we consider. Since $N^\mathrm{tot}_\mathrm{EMRI}$ depends weakly on the mass of the orbiter at fixed $M_\bullet$, $\langle N_\mathrm{QPE} \rangle$ follows the same qualitative trends as $\mu^{T,e,i}$. Similarly, the scaling of loss cone rates with $M_\bullet$ mildly affects that of $\mu^{T,e,i}$.

Finally, we combine our results with the MBH mass function to compute the cosmic QPE abundance $\langle n_\mathrm{QPE} \rangle$, as described in Subsection~\ref{subsec:active_qpes}.
In Fig.~\ref{fig:qpe_vol} we present $\langle n_\mathrm{QPE} \rangle$ for $T_\mathrm{QPE} = \{2,9,20,48\}$ h and $e^\mathrm{max}_\mathrm{QPE}=\{0.05, 0.2, 1\}$ \footnote{Physically, the maximum possible eccentricity is set by the minimum pericenter $r_\mathrm{cap}$ or $r_t$.}. Considering $T_\mathrm{QPE}\gtrsim9$h, rates can vary by  $2-3$ orders of magnitude depending on the adopted
eccentricity constraints, and within an order of magnitude depending on the period constraints. Overall, stellar QPEs are $\sim10^3$ more numerous than sBHs QPEs at fixed constraints.

\citet{2024Arcodia} estimate a cosmic QPE abundance of $(1.7\times10^{-7}-5\times10^{-6})$ Mpc$^{-3}$, which is compatible with our estimates for QPEs in the disk-impact model from stellar EMRIs. By removing eccentricity and inclination constraints, sBHs QPEs are compatible with the lower end of such estimates. See the next section for a more detailed discussion.

\section{Discussion} \label{sec:discussion}

We will now discuss the assumptions and the limitations of our results.

\subsection{EMRI formation channels}
Our model focuses on the classical EMRI formation channel, but various alternative dynamical channels have been proposed. Among those, some might have cosmological rates comparable to or larger than spherical two-body relaxation, especially when restricting to small eccentricities and sBHs. For example, a member of a stellar binary might be deposited as the result of a three-body encounter with the central MBH \citep[Binary deposition][]{2005Miller}. Such a channel might favour small eccentricity with comparable rates \citep[][]{2021Raveh}.
If the central MBH is surrounded by a disk, it can capture compact objects because of inclination damping, and further reduce the orbital eccentricity as the compact object migrates toward the MBH \citep[e.g.,][]{1991Syer,1999Subr,2001Karas,Levin2007,Kocsis2011} . This channel seems to be very promising for producing EMRIs (see \citet{2021Pan}) with small eccentricity \citep{2025Sun}. Alternatively, a star or an sBH might form in the disk \citep[e.g.,][]{1993Artymowicz,2003Levin,2004Goodman}, or multiple mergers can increase the size of sBHs. Since EMRI formation rates in these scenarios suffer of very large uncertainties, we do not venture into computing the corrisponding expected QPE rates; however, the current formalism can be readily applied to any EMRI formation channel.

\begin{figure*}
    \centering

    \includegraphics[width=\textwidth]{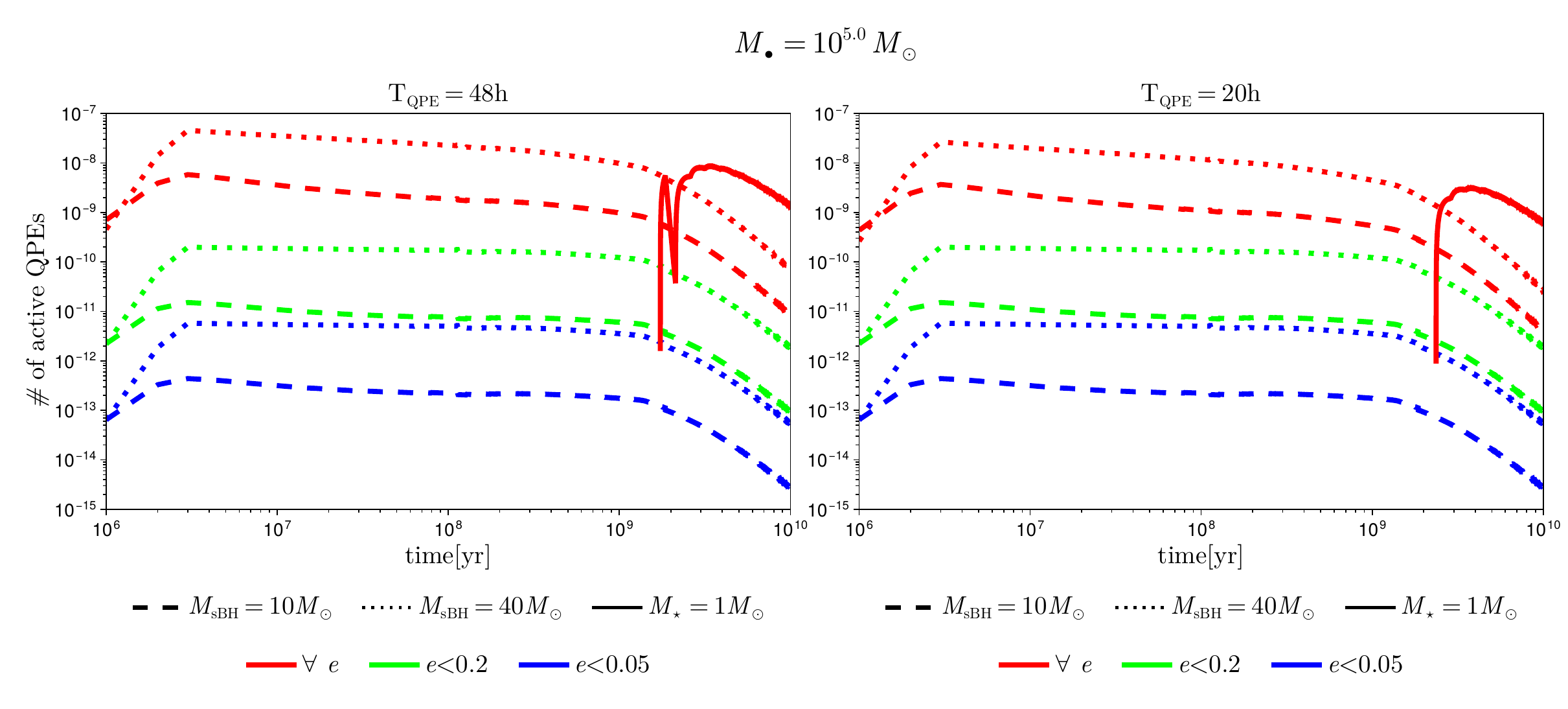}
   
    \vspace{0.2cm} 
    \includegraphics[width=\textwidth]{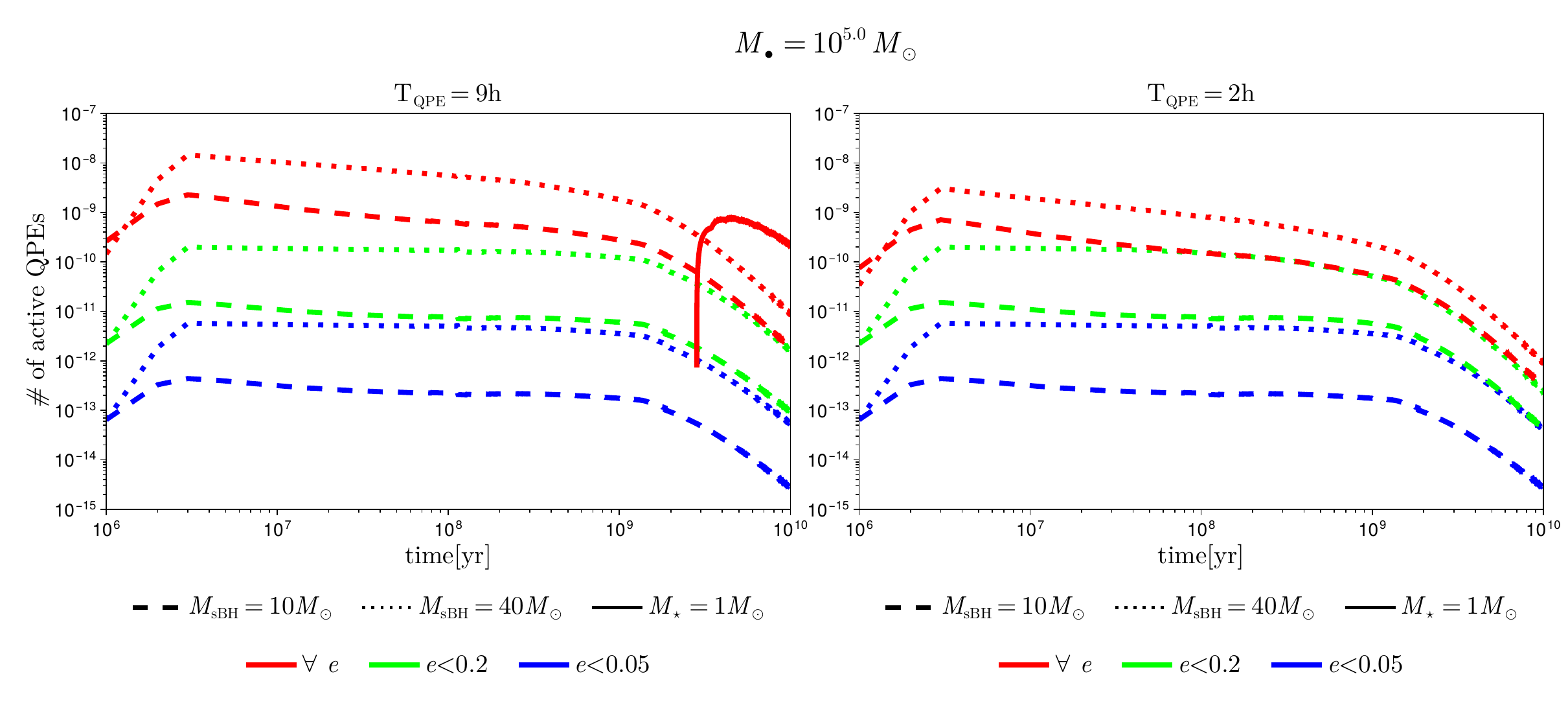}
    \captionsetup{justification=centering} 
    \caption{Number of active QPEs of sBHs of $10$M$_\odot$ (dashed line), $40$M$_\odot$ (dotted line), and stars of $1$M$_\odot$ (solid line) around a MBH of $10^5$M$_\odot$. Different colors are for different limiting eccentricities, as marked in the figure.  The four panels are for different values of $T_{\rm QPE}$ (on the first row: $T_\mathrm{QPE}$=48h on the left, 20h on the right, on the second row: $T_\mathrm{QPE}$=9h on the left, 2h on the right).}
    \label{fig:QPE_rates_1}
\end{figure*}

\begin{figure*}
    \centering

    \includegraphics[width=\textwidth]{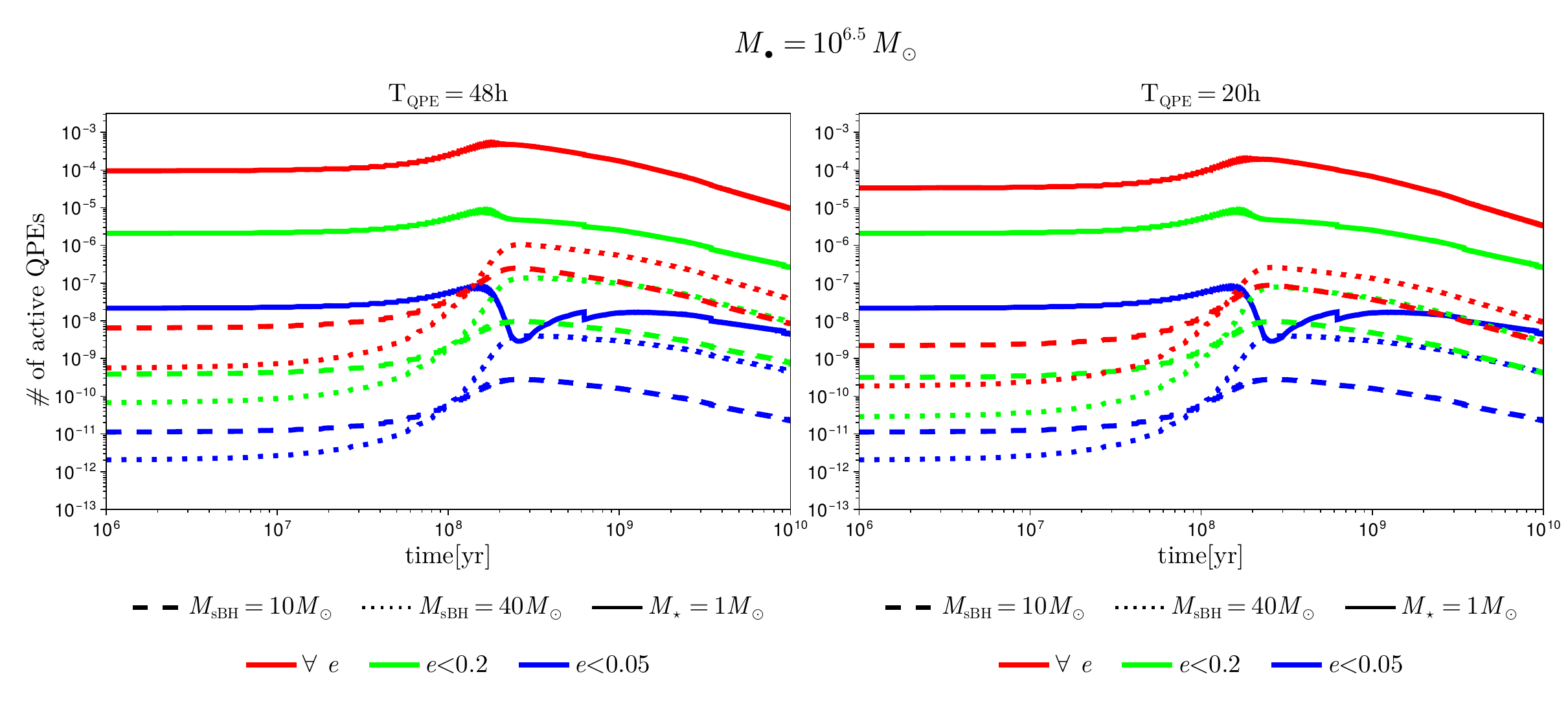}
   
    \vspace{0.2cm} 
    \includegraphics[width=\textwidth]{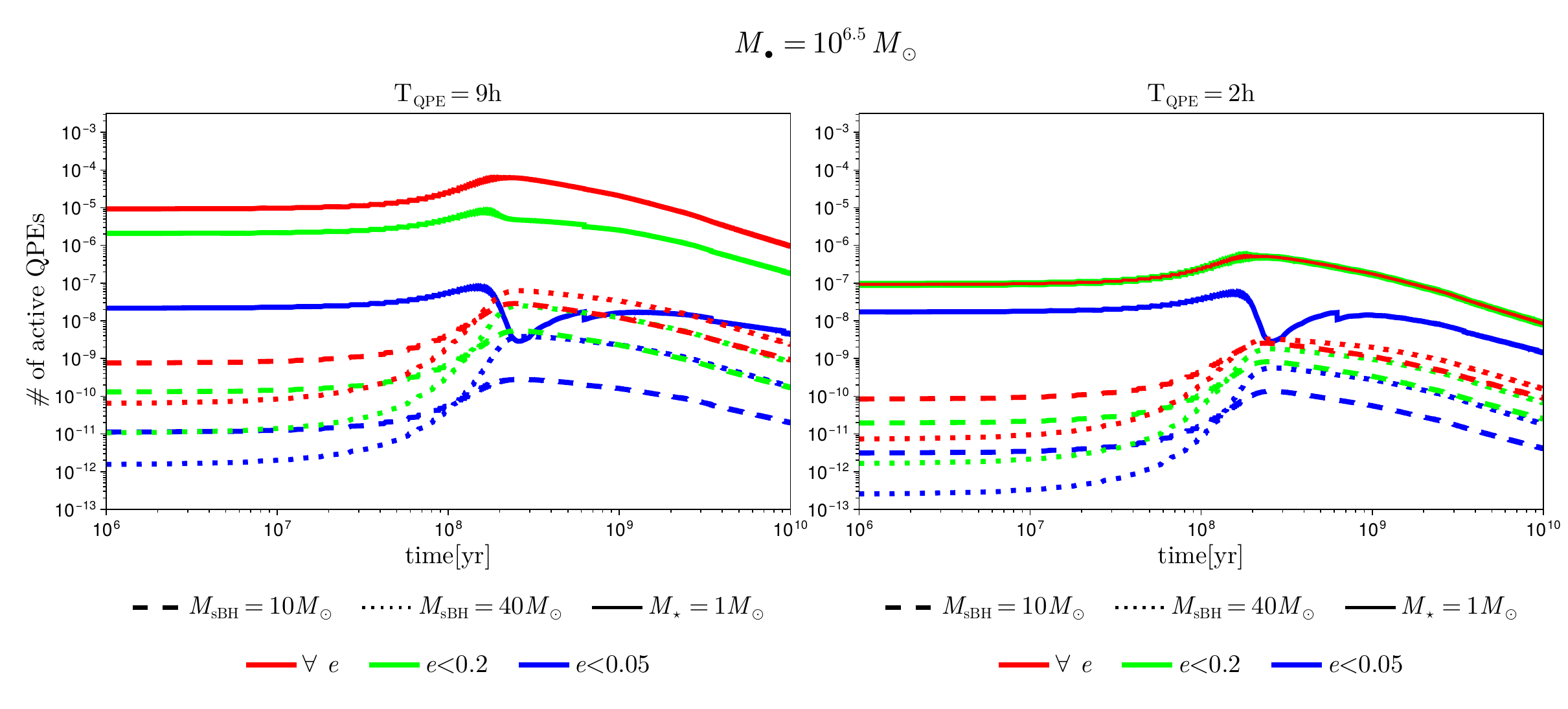}
    \captionsetup{justification=centering} 

    \caption{Same as Fig.~\ref{fig:QPE_rates_1} but for a central MBH of $3\times 10^6$M$_\odot$}
    \label{fig:QPE_rates_2}
\end{figure*}

\begin{figure*}
    \centering
    \includegraphics[width=0.8\linewidth]{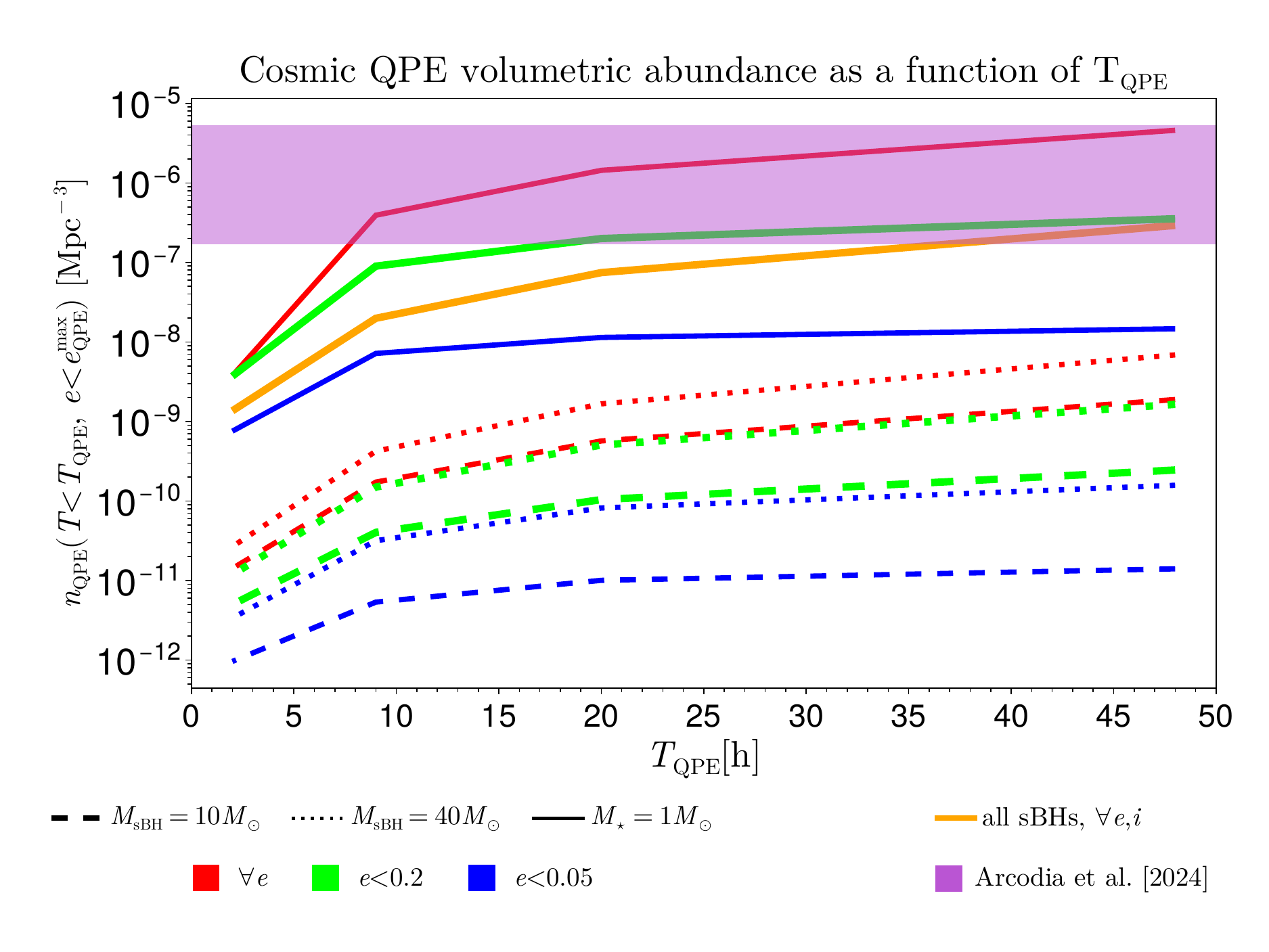}
    \caption{Cosmic volumetric abundance assuming a limiting eccentricity of 0.05, 0.2, 1.0 for sBHs of $10$M$_\odot$ (dashed line) and $40$M$_\odot$ (dotted line). The cosmic volumetric abundance of stars (solid line) is portrayed at e=0.05, 0.2, 1-$r_t/a_\mathrm{QPE}$. The orange solid line corresponds to the sum of the $n_\mathrm{QPE}$ of both sBHs without accounting for the constraints on $e$ and $i$. The violet band corresponds to the QPE volumetric abundance found by \citet{2024Arcodia}.}
    \label{fig:qpe_vol}
\end{figure*}

\subsection{Dependence of the cosmic volumetric abundance on the TDE disk lifetime} \label{subsec:disk_lifetime_volumetric} 

Throughout this work we adopt a fiducial TDE disk lifetime \(\tau_{\rm disk}=10\,{\rm yr}\), motivated by viscous disk-evolution models for TDE debris disks \citep[e.g.,][]{1990Cannizzo,2011Montesinos_Armijo,2024Piro}. The physical value of \(\tau_{\rm disk}\) is uncertain and may depend on several parameters, including the mass of the disrupted star, the MBH mass, the accretion efficiency, the viscosity parameter, the disk aspect ratio, and the circularization radius. Therefore, the assumption of a single disk lifetime for all MBH masses should be regarded as an effective choice rather than as a universal property of TDE disks. The impact of this assumption on our results can be read directly from the definition of the active QPE abundance $N_{\rm QPE}$, which is linearly proportional to \(\tau_{\rm disk}\)  (see Eq.~\eqref{eq:active_qpe}). This linear dependence then propagates directly to the cosmic volumetric abundance, since \(n^i_{\rm QPE}\) is obtained by multiplying \(N_{\rm QPE}(M_i,t)\) by the MBH mass function in each mass bin (Eq.~\eqref{eq:nqpe_bin}), and \(\langle n_{\rm QPE}\rangle\) is obtained by time-averaging and summing over the MBH mass bins (Eqs.~\eqref{eq:nqpe_bin_avg}--\eqref{eq:nqpe_total_avg}). Consequently, if a different but mass-independent value of $\tau_{\rm disk}$ is adopted, all volumetric abundances shown in Fig.~\ref{fig:qpe_vol} can be rescaled by the same multiplicative factor. If instead $\tau_{\rm disk}$ depends on $M_\bullet$, the rescaling must be performed separately in each MBH mass bin, potentially altering the relative contribution of the different bins to the total abundance. 
Such a rescaling is sufficient only when the underlying stellar and dynamical model is kept fixed. This alters the range of MBH masses for which disruption occurs, as well as the resulting disk properties, potentially affecting the TDE rate, disk lifetime, and QPE-selection factor. Therefore, varying only $\tau_{\rm disk}$ produces a simple multiplicative rescaling, whereas changing the stellar model may require the TDE and QPE-selection calculations to be recomputed consistently and can modify both the normalization and the MBH-mass dependence of the predicted QPE abundance. Our fiducial comparison with the observational abundance estimated by \citet{2024Arcodia} is consistent because both analyses adopt a reference timescale of $10{\rm yr}$. Although the two timescales have different physical meanings, our disk lifetime coincides with the fiducial QPE lifetime, $\tau_{\rm life}=10\mathrm{yr}$, adopted by \citet{2024Arcodia} when converting their observed abundance into a formation rate. However, their abundance estimate is independent of $\tau_{\rm life}$, whereas our predicted abundance scales linearly with $\tau_{\rm disk}$. Significantly different values of $\tau_{\rm disk}$ would therefore shift our predictions and modify the comparison.

\subsection{Dynamics in the NSC}

As we explained in Sec.~\ref{sec:model}, the effects of GW emission are not included in 1D Fokker-Planck models as \textsc{PhaseFlow}, primarily for the significant impact of eccentricity. We neglect the spin of the primary MBH, that might increase the number of EMRIs and affect the inclination distribution \citep[e.g.,][]{2025Zhang, 2025Sun}.
Deviations from spherical symmetry might increase loss cone rates (most importantly the TDE rate), even though the impact might be minor for the small MBHs we consider \citep{2013Vasiliev}.
Similarly, a steep gaseous profile might affect the evolution of objects because of significant dynamical friction \citep{2025Rozner}.

Currently, loss cone theory neglects partial tidal disruptions and Roche lobe overflows of the star, accounting solely for the possibility of complete tidal disruption \citep{1978Cohn,2013Merritt, 2025Broggi}. The TDE rate might be marginally affected for small $M_\bullet$ \citep[e.g.,][]{2023Bortolas, 2024Broggi}. However, as EMRIs require repeated passages at small pericenter, these effects might significantly impact the rate of stellar EMRIs and, consequently, the related QPE abundance.

\subsection{Galactic population}

We assume that every MBH is surrounded by a NSC with properties fixed by its mass. Naturally, not all MBHs have a NSC. Moreover,
the properties of NSCs vary systematically with host galaxy mass and morphology \citep{2020Neumayer}, suggesting that the initial phase of the relaxation process and the total available mass budget in our simple NSC model might be inadequate. At later stage, relaxation dictates the stellar profile, possibly reducing the uncertainty when considering the cosmic population.
Finally, we represent the stellar population using only three discrete mass components. This approximation captures the main effects of mass segregation, but it necessarily simplifies the underlying continuous stellar mass spectrum. Different stellar masses segregate on different timescales and contribute differently to relaxation. In particular, massive remnants tend to concentrate more strongly toward the MBH and can enhance the relaxation experienced by lighter objects, while low-mass stars remain less centrally concentrated. A broader mass spectrum could therefore modify both the relative abundance of the different populations near the MBH and the time-dependent TDE and EMRI rates \citep{2022Bortolas}. Moreover, the mass of the inspiraling object directly affects its GW-driven evolution and coalescence timescale. Therefore, if timing models provide observational constraints on the masses of stellar EMRIs, a more detailed treatment of the population mass spectrum would be required to model both their dynamical distribution and their GW evolution consistently. Our three-component treatment should thus be regarded as an effective description rather than a complete representation of the mass dependence of the population.

\subsection{Observability of QPEs}
We do not include observability constraints on disk/EMRI systems. However, emission mechanisms with a preferential axis or the properties of the TDE disk might reduce the number of observable systems.
In fact, electromagnetic emission in the disk-impact model is not fully understood. The bubble generated as the object crosses the disk cools until it becomes optically thin; at that stage, its temperature settles to the inferred temperature $T_{\mathrm{obs}}$, crucially set by the cooling process.
These open points  directly impact our rates through orbital constraints. For example, very eccentric EMRIs from sBHs might reach pericenters so small that the density of the disk is considerable; despite the small Bondi radius, the passage might strip as much mass as in the circular case, producing a comparably luminous flare, thus relaxing eccentricity constraints.

\subsection{Constraints from timing models} The first results from fitting the timing of QPEs with dynamical models \citep[see e.g.,][]{2024PhRvD.110h3019Z} indicate a preference for small eccentricity, with no assumptions on the type of orbiter. While this might be caused by the size of Bondi radius for sBHs, stellar QPEs from standard two-body relaxation should favour large eccentricity. Restricting the maximum eccentricity affects expectations on the number density of QPEs by orders of magnitude, possibly pushing it outside of currently inferred values.

\section{Conclusions} \label{sec:conclusions}

We built a framework to compute the number of expected QPEs produced by two-body relaxation in the disk-impact model. We assume a Schwarzschild MBH and a spherically symmetric potential.

We linked the QPE volumetric abundance to the formation rate of EMRIs and TDEs, which we modeled by simulating the relaxation of NSCs around MBHs of mass $M_\bullet = 10^5-10^8 M_\odot$ using the Fokker-Planck integration code \textsc{PhaseFlow} \citep{2017Vasiliev}.

Each NSC was modeled with three mass components: main sequence stars of $1$M$_\odot$, and two families of sBHs with masses of $10$M$_\odot$ and $40$M$_\odot$. We evolved the systems for $10$Gyr,
tracking the evolution of the TDE and EMRI rates. Finally, we selected EMRI events within the constraints arising from the emission model: the maximum orbital period $T_\mathrm{QPE}$, and for sBHs the maximum eccentricity $e^\mathrm{max}_\mathrm{QPE} \simeq 0.2$ and the maximum inclination $0 < i < i^\mathrm{max}_\mathrm{QPE}\simeq20^\circ$. By assuming that all MBHs are surrounded by an NSC, we combined our results with the BH mass function to estimate the QPE abundance in the universe.

We found that the QPE volumetric abundance (or number density) can vary in the range $10^{-12} \rm Mpc^{-3}$ to $10^{-6} \rm Mpc^{-3}$ depending on the imposed constraints (see Fig.~\ref{fig:qpe_vol}).
QPEs powered by impacting stars can be as frequent as suggested by first observational estimates \citep{2024Arcodia}.
Conversely, the constraints on the orbital properties of sBHs significantly reduce the number of observable QPEs, which are $\sim 10^3$ times less numerous than stellar ones and generally below observational estimates.

Removing inclination and eccentricity constraints, the total number of QPEs from sBHs is compatible to the lower end of observational predictions for $T_\mathrm{QPE}=48$h, whereas the number of QPEs from stars is compatible to observational estimates for $T_\mathrm{QPE}>9$h.
In general, two-body relaxation induces a preference for eccentric sources, in contrast with the picture emerging from fitting of QPE timing and, for sBH-driven QPEs, from constraints on the Bondi radius. Restricting to low-eccentricity sources significantly reduces the rate, possibly to values incompatible with observational estimates both for stars and sBHs. In our model, restricting sBH-driven QPEs to low-inclination systems reduces their predicted rate by a factor of $\sim 30$. By contrast, \citet{2026Liu} adopt the Hill radius as the relevant impact scale for sBHs, which is independent of orbital inclination. This assumption therefore allows a broader range of orbital configurations and increases the predicted number of sBH-driven QPEs.

By including other formation channels for EMRIs, a more detailed dynamical model and more refined galactic models, it might be possible to probe, constrain and rule-out dynamical models of QPEs. Shedding light on their nature and understanding their possible connection to EMRIs might unveil them as an electromagnetic counterpart of future LISA sources, opening new avenues in multimessenger astronomy.

\begin{acknowledgements}
 AS acknowledges financial support provided under the European Union’s H2020 ERC Advanced Grant ``PINGU'' (Grant Agreement: 101142079). MB acknowledges support from the Italian Ministry for Universities and Research (MUR) program “Dipartimenti di Eccellenza 2023-2027”, within the framework of the activities of the Centro Bicocca di Cosmologia Quantitativa (BiCoQ). We thank A. Franchini for useful discussions and valuable comments. C.M.A. and L.B. thank D. Mancieri for helpful suggestions, especially on the figures and their graphical presentation.       
\end{acknowledgements}

\bibliographystyle{aa} 
\bibliography{references}

\begin{landscape}
\begin{appendix} 
\section{Additional tables on $\mu^\mathrm{T,e,i}$ and $\langle N_\mathrm{QPE} \rangle$} \label{app:tables}
\thispagestyle{empty} % remove the default (vertical) page number
\LandscapePageNumber  % add the horizontal one

\begin{table}[h!]
\centering
\small

\setlength{\tabcolsep}{2.5pt} % tighter spacing
\hspace*{-2cm} % shift left
\resizebox{1.1\linewidth}{!}{%
\begin{tabular}{c|c|
c|c|c|c|c|c|c|c|c|c|c|c|c|c|c|c|c|c|c|c|c}
\hline

% ----------------- Top label: SMBH masses -----------------
& & \multicolumn{21}{c}{$\boldsymbol{M_\bullet}$} \\
\hline

% ----------------- SMBH mass headers -----------------
& &
\multicolumn{3}{c|}{$10^5\,M_\odot$} &
\multicolumn{3}{c|}{$3\times10^5\,M_\odot$} &
\multicolumn{3}{c|}{$10^6\,M_\odot$} &
\multicolumn{3}{c|}{$3\times10^6\,M_\odot$} &
\multicolumn{3}{c|}{$10^7\,M_\odot$} &
\multicolumn{3}{c|}{$3\times10^7\,M_\odot$} &
\multicolumn{3}{c}{$10^8\,M_\odot$} \\
\hline

% ----------------- CO mass label row -----------------
& &
\multicolumn{3}{c|}{$\boldsymbol{M_{\rm CO}}$} &
\multicolumn{3}{c|}{$\boldsymbol{M_{\rm CO}}$} &
\multicolumn{3}{c|}{$\boldsymbol{M_{\rm CO}}$} &
\multicolumn{3}{c|}{$\boldsymbol{M_{\rm CO}}$} &
\multicolumn{3}{c|}{$\boldsymbol{M_{\rm CO}}$} &
\multicolumn{3}{c|}{$\boldsymbol{M_{\rm CO}}$} &
\multicolumn{3}{c}{$\boldsymbol{M_{\rm CO}}$} \\
\hline

% ----------------- CO mass entries -----------------
& &
\textit{10 $M_\odot$} & \textit{40 $M_\odot$} & \textit{1 $M_\odot$} &
\textit{10 $M_\odot$} & \textit{40 $M_\odot$} & \textit{1 $M_\odot$} &
\textit{10 $M_\odot$} & \textit{40 $M_\odot$} & \textit{1 $M_\odot$} &
\textit{10 $M_\odot$} & \textit{40 $M_\odot$} & \textit{1 $M_\odot$} &
\textit{10 $M_\odot$} & \textit{40 $M_\odot$} & \textit{1 $M_\odot$} &
\textit{10 $M_\odot$} & \textit{40 $M_\odot$} & \textit{1 $M_\odot$} &
\textit{10 $M_\odot$} & \textit{40 $M_\odot$} & \textit{1 $M_\odot$} \\
\hline

% ----------------- Row labels (T_QPE, e) -----------------
$\boldsymbol{T_{\mathrm{QPE}}}$ & $\boldsymbol{e_{\mathrm{QPE}}^\mathrm{max}}$ &
\multicolumn{21}{c}{} \\ 
\hline

%rows
 & 0.05 & $1.6\times10^{-6}$ & $3.9\times10^{-6}$ & 0.0 & $2.3\times10^{-6}$ & $5.7\times10^{-6}$ & 0.0 & $3.3\times10^{-6}$ & $8.2\times10^{-6}$ & $3.1\times10^{-6}$ & $4.8\times10^{-6}$ & $1.2\times10^{-5}$ & $6.1\times10^{-5}$ & $6.7\times10^{-6}$ & $1.6\times10^{-5}$ & $1.4\times10^{-4}$ & $8.8\times10^{-6}$ & $1.4\times10^{-5}$ & $2.52\times10^{-4}$ & $1.3\times10^{-5}$ & $6.4\times10^{-6}$ & $4.4\times10^{-4}$ \\
48\,h & 0.2 & $5.6\times10^{-5}$ & $1.3\times10^{-4}$ & 0.0 & $8.1\times10^{-5}$ & $2.0\times10^{-4}$ & 0.0012 & $1.1\times10^{-4}$ & $2.8\times10^{-4}$ & 0.0035 & $1.6\times10^{-4}$ & $3.0\times10^{-4}$ & 0.0051 & $1.7\times10^{-4}$ & $1.8\times10^{-4}$ & 0.0061 & $1.0\times10^{-4}$ & $7.1\times10^{-5}$ & 0.0073 & $7.7\times10^{-5}$ & $3.6\times10^{-5}$ & 0.0036 \\
 & 1 & 0.0074 & 0.0083 &  1.0 & 0.0057 & 0.0056 & 0.75 & 0.0040 & 0.0031 & 0.47 & 0.0024 & 0.0014 & 0.26 & 0.0012 & $5.5\times10^{-4}$ & 0.13 & $4.6\times10^{-4}$ & $1.6\times10^{-4}$ & 0.056 & $3.2\times10^{-4}$ & $1.5\times10^{-4}$ & 0.013 \\
\hline

 & 0.05 & $1.6\times10^{-6}$ & $3.9\times10^{-6}$ & 0.0 & $2.3\times10^{-6}$ & $5.7\times10^{-6}$ & 0.0 & $3.3\times10^{-6}$ & $8.2\times10^{-6}$ & $3.1\times10^{-6}$ & $4.8\times10^{-6}$ & $1.1\times10^{-5}$ & $6.1\times10^{-5}$ & $6.5\times10^{-6}$ & $9.5\times10^{-6}$ & $1.4\times10^{-4}$ & $5.0\times10^{-6}$ & $4.4\times10^{-6}$ & $2.47\times10^{-4}$ & $4.3\times10^{-6}$ & $2.0\times10^{-6}$ & $1.9\times10^{-4}$ \\
20\,h & 0.2 & $5.6\times10^{-5}$ & $1.3\times10^{-4}$ & 0.0 & $8.1\times10^{-5}$ & $1.9\times10^{-4}$ & 0.0012 & $1.1\times10^{-4}$ & $1.9\times10^{-4}$ & 0.0035 & $1.1\times10^{-4}$ & $1.1\times10^{-4}$ & 0.0051 & $7.1\times10^{-5}$ & $4.6\times10^{-5}$ & 0.0052 & $3.4\times10^{-5}$ & $1.5\times10^{-5}$ & 0.0032 & $2.3\times10^{-5}$ & $1.1\times10^{-5}$ & 0.0011 \\
 & 1 & 0.0039 & 0.0036 & 0.49 & 0.0026 & 0.0020 & 0.34 & 0.0015 & $9.2\times10^{-4}$ & 0.20 & $8.1\times10^{-4}$ & $3.6\times10^{-4}$ & 0.099 & $3.5\times10^{-4}$ & $1.2\times10^{-4}$ & 0.041 & $1.1\times10^{-4}$ & $2.9\times10^{-5}$ & 0.015 & $7.4\times10^{-5}$ & $3.3\times10^{-5}$ & 0.0025 \\
\hline

 & 0.05 & $1.6\times10^{-6}$ & $3.9\times 10^{-6}$ & 0.0 & $2.3\times10^{-6}$ & $5.7\times 10^{-6}$ & 0.0 & $3.3\times10^{-6}$ & $7.9\times 10^{-6}$ & $3.1\times 10^{-6}$ & $4.5\times10^{-6}$ & $6.4\times 10^{-6}$ & $6.1\times 10^{-5}$ & $3.8\times10^{-6}$ & $3.2\times 10^{-6}$ & $1.4\times 10^{-4}$ & $2.0\times10^{-6}$ & $1.2\times 10^{-6}$ & $1.5\times10^{-4}$ & - & - & - \\
9\,h & 0.2 & $5.6\times 10^{-5}$ & $1.3\times 10^{-4}$ & 0.0 & $7.9\times 10^{-5}$ & $1.3\times 10^{-4}$ & 0.0012 & $7.6\times 10^{-5}$ & $7.2\times 10^{-5}$ & 0.0035 & $5.0\times 10^{-5}$ & $3.2\times 10^{-5}$ & 0.0044 & $2.6\times 10^{-5}$ & $1.2\times 10^{-5}$ & 0.0025 & $1.1\times 10^{-5}$ & $3.3\times 10^{-6}$ & 0.0012 & - & - & - \\
 & 1 & 0.0019 & 0.0014 & 0.15 & 0.0011 & $6.7\times10^{-4}$ & 0.12 & $5.8\times10^{-4}$ & $2.6\times 10^{-4}$ & 0.067 & $2.7\times 10^{-4}$ & $9.2\times 10^{-5}$ & 0.030 & $9.9\times 10^{-5}$ & $2.6\times 10^{-5}$ & 0.011 & $2.5\times 10^{-5}$ & $5.2\times 10^{-6}$ & 0.0038 & - & - & - \\
\hline

 & 0.05 & $1.6\times10^{-6}$ & $3.7\times10^{-6}$ & 0.0 & $2.2\times10^{-6}$ & $3.2\times10^{-6}$ & 0.0 & $1.9\times10^{-6}$ & $1.7\times10^{-6}$ & $3.1\times10^{-6}$ & $1.2\times10^{-6}$ & $7.2\times10^{-7}$ & $2.9\times10^{-5}$ & $6.2\times10^{-7}$ & $2.6\times10^{-7}$ & $2.1\times10^{-5}$ & - & - & - & - & - & - \\
2\,h & 0.2 & $3.9\times10^{-5}$ & $3.9\times10^{-5}$ & 0.0 & $2.7\times10^{-5}$ & $1.9\times10^{-5}$ & $3.5\times10^{-4}$ & $1.5\times10^{-5}$ & $7.3\times10^{-6}$ & $4.4\times10^{-4}$ & $7.4\times10^{-6}$ & $2.6\times10^{-6}$ & $2.5\times10^{-4}$ & $3.0\times10^{-6}$ & $7.3\times10^{-7}$ & $1.0\times10^{-4}$ & - & - & - & - & - & - \\
 & 1 & $3.7\times10^{-4}$ & $1.7\times10^{-4}$ & 0.0 & $1.8\times10^{-4}$ & $6.4\times10^{-4}$ & $3.8\times10^{-4}$ & $7.4\times10^{-5}$ & $2.1\times10^{-5}$ & $4.5\times10^{-4}$ & $2.6\times10^{-5}$ & $5.7\times10^{-6}$ & $2.6\times10^{-4}$ & $6.0\times10^{-6}$ & $1.1\times10^{-6}$ & $1.0\times10^{-4}$ & - & - & - & - & - & - \\
\hline

\end{tabular}
}
\caption{QPE selection factor $\mu^{T,e,i} = N_{CO} / N^{tot}_\mathrm{EMRI}$ for different values of the maximum period $T_\mathrm{QPE}$, and the maximum eccentricity $e^\mathrm{max}_\mathrm{QPE}$ compatible with QPEs. We consider NSCs with central MBH of mass $M_\bullet$ and EMRIs from COs of different mass $M_{\mathrm{CO}}$. For sBHs ($M_\mathrm{CO}=10, 40\, M_\odot$) $\mu^{T,e,i}$ also accounts for inclination constraints $0^\circ <i <  20^\circ$. ``-'' means that the period at $r_\mathrm{cap}$ is longer than $T_\mathrm{QPE}$. }
\label{table:1}
\end{table}

\begin{table}[h!]
\centering
\small

\setlength{\tabcolsep}{2.5pt} 
\hspace*{-2cm} 
\resizebox{1.08\linewidth}{!}{
\begin{tabular}{c|c|
c|c|c|c|c|c|c|c|c|c|c|c|c|c|c|c|c|c|c|c|c}
\hline

% ----------------- Top label: SMBH masses -----------------
& & \multicolumn{21}{c}{$\boldsymbol{M_\bullet}$} \\
\hline

% ----------------- SMBH mass headers -----------------
& &
\multicolumn{3}{c|}{$10^5\,M_\odot$} &
\multicolumn{3}{c|}{$3\times10^5\,M_\odot$} &
\multicolumn{3}{c|}{$10^6\,M_\odot$} &
\multicolumn{3}{c|}{$3\times10^6\,M_\odot$} &
\multicolumn{3}{c|}{$10^7\,M_\odot$} &
\multicolumn{3}{c|}{$3\times10^7\,M_\odot$} &
\multicolumn{3}{c}{$10^8\,M_\odot$} \\
\hline

% ----------------- CO mass label row -----------------
& &
\multicolumn{3}{c|}{$\boldsymbol{M_{\rm CO}}$} &
\multicolumn{3}{c|}{$\boldsymbol{M_{\rm CO}}$} &
\multicolumn{3}{c|}{$\boldsymbol{M_{\rm CO}}$} &
\multicolumn{3}{c|}{$\boldsymbol{M_{\rm CO}}$} &
\multicolumn{3}{c|}{$\boldsymbol{M_{\rm CO}}$} &
\multicolumn{3}{c|}{$\boldsymbol{M_{\rm CO}}$} &
\multicolumn{3}{c}{$\boldsymbol{M_{\rm CO}}$} \\
\hline

% ----------------- CO mass entries -----------------
& &
\textit{10 $M_\odot$} & \textit{40 $M_\odot$} & \textit{1 $M_\odot$} &
\textit{10 $M_\odot$} & \textit{40 $M_\odot$} & \textit{1 $M_\odot$} &
\textit{10 $M_\odot$} & \textit{40 $M_\odot$} & \textit{1 $M_\odot$} &
\textit{10 $M_\odot$} & \textit{40 $M_\odot$} & \textit{1 $M_\odot$} &
\textit{10 $M_\odot$} & \textit{40 $M_\odot$} & \textit{1 $M_\odot$} &
\textit{10 $M_\odot$} & \textit{40 $M_\odot$} & \textit{1 $M_\odot$} &
\textit{10 $M_\odot$} & \textit{40 $M_\odot$} & \textit{1 $M_\odot$} \\
\hline

% ----------------- Row labels (T_QPE, e) -----------------
$\boldsymbol{T_{\mathrm{QPE}}}$ & $\boldsymbol{e_{\mathrm{QPE}}^\mathrm{max}}$ &
\multicolumn{21}{c}{} \\ % optional blank row for spacing
\hline

% ----------------- rows -----------------
 & 0.05 & $5.0\times10^{-14}$ & $10^{-12}$ & 0.0 & $6.0\times10^{-13}$ & $1.2\times10^{-11}$ & 0.0 & $6.6\times10^{-12}$ & $1.3\times10^{-10}$ & $1.1\times10^{-11}$ & $6.6\times10^{-11}$ & $1.2\times10^{-9}$ & $10^{-10}$ & $8.6\times10^{-10}$ & $1.4\times10^{-8}$ & $4.2\times10^{-7}$ & $4.3\times10^{-9}$ & $3.4\times10^{-8}$ & $3.7\times10^{-6}$ & $6.7\times10^{-10}$ & $2.3\times10^{-10}$ & $7.4\times10^{-6}$ \\
48\,h & 0.2 & $1.7\times10^{-12}$ & $3.5\times10^{-11}$ & 0.0 & $2.1\times10^{-11}$ & $4.2\times10^{-10}$ & $1.8\times10^{-10}$ & $2.3\times10^{-10}$ & $4.5\times10^{-9}$ & $3.6\times10^{-8}$ & $2.3\times10^{-9}$ & $3.5\times10^{-8}$ & $10^{-6}$ & $2.2\times10^{-8}$ & $1.6\times10^{-7}$ & $1.8\times10^{-5}$ & $5.0\times10^{-8}$ & $1.7\times10^{-7}$ & $10^{-4}$ & $4.0\times10^{-9}$ & $1.3\times10^{-9}$ & $6.0\times10^{-5}$ \\
 & 1 & $2.8\times10^{-10}$ & $2.9\times10^{-9}$ & $3.5\times10^{-9}$ & $1.9\times10^{-9}$ & $1.6\times10^{-8}$ & $4.0\times10^{-7}$ & $9.5\times10^{-9}$ & $6.5\times10^{-8}$ & $7.4\times10^{-6}$ & $3.8\times10^{-8}$ & $1.8\times10^{-7}$ & $6.1\times10^{-5}$ & $1.5\times10^{-7}$ & $4.9\times10^{-7}$ & $3.8\times10^{-4}$ & $2.2\times10^{-7}$ & $3.4\times10^{-7}$ & $8.2\times10^{-4}$ & $1.6\times10^{-8}$ & $5.4\times10^{-9}$ & $2.2\times10^{-4}$ \\
\hline

 & 0.05 & $5.0\times10^{-14}$ & $10^{-12}$ & 0.0 & $6.0\times10^{-13}$ & $1.2\times10^{-11}$ & 0.0 & $6.6\times10^{-12}$ & $1.3\times10^{-10}$ & $1.1\times10^{-11}$ & $6.6\times10^{-11}$ & $1.2\times10^{-10}$ & $9.9\times10^{-9}$ & $8.2\times10^{-10}$ & $8.4\times10^{-9}$ & $4.2\times10^{-7}$ & $2.4\times10^{-9}$ & $1.1\times10^{-8}$ & $3.6\times10^{-6}$ & $2.2\times10^{-10}$ & $7.3\times10^{-11}$ & $3.1\times10^{-6}$ \\
20\,h & 0.2 & $1.7\times10^{-12}$ & $3.5\times10^{-11}$ & 0.0 & $2.1\times10^{-11}$ & $4.2\times10^{-10}$ & $1.8\times10^{-10}$ & $2.2\times10^{-10}$ & $3.5\times10^{-9}$ & $3.6\times10^{-8}$ & $1.7\times10^{-9}$ & $1.3\times10^{-8}$ & $10^{-6}$ & $9.1\times10^{-9}$ & $4.1\times10^{-8}$ & $1.6\times10^{-5}$ & $1.6\times10^{-8}$ & $3.7\times10^{-8}$ & $4.7\times10^{-5}$ & $1.2\times10^{-9}$ & $4.0\times10^{-10}$ & $1.9\times10^{-5}$ \\
 & 1 & $1.6\times10^{-10}$ & $1.3\times10^{-9}$ & $1.3\times10^{-9}$ & $9.0\times10^{-10}$ & $6.2\times10^{-9}$ & $1.8\times10^{-7}$ & $3.9\times10^{-9}$ & $1.9\times10^{-8}$ & $3.2\times10^{-6}$ & $1.3\times10^{-8}$ & $4.5\times10^{-8}$ & $2.3\times10^{-5}$ & $4.4\times10^{-8}$ & $1.0\times10^{-7}$ & $1.2\times10^{-4}$ & $5.3\times10^{-8}$ & $7.0\times10^{-8}$ & $2.2\times10^{-4}$ & $3.8\times10^{-9}$ & $1.2\times10^{-9}$ & $4.3\times10^{-5}$ \\
\hline

 & 0.05 & $5.0\times10^{-14}$ & $10^{-12}$ & 0.0 & $6.0\times10^{-13}$ & $1.2\times 10^{-11}$ & 0.0 & $6.6\times10^{-12}$ & $1.3\times 10^{-10}$ & $1.1\times 10^{-11}$ & $6.4\times10^{-11}$ & $7.7\times 10^{-10}$ & $9.9\times 10^{-9}$ & $4.8\times10^{-10}$ & $2.8\times 10^{-9}$ & $4.2\times 10^{-7}$ & $9.8\times10^{-10}$ & $2.9\times 10^{-9}$ & $2.1\times10^{-6}$ & - & - & - \\
9\,h & 0.2 & $1.7\times 10^{-12}$ & $3.5\times 10^{-11}$ & 0.0 & $2.1\times 10^{-11}$ & $3.4\times 10^{-10}$ & $1.8\times10^{-10}$ & $1.8\times 10^{-10}$ & $1.5\times 10^{-9}$ & $3.6\times10^{-8}$ & $7.9\times 10^{-10}$ & $4.0\times 10^{-9}$ & $9.4\times10^{-7}$ & $3.3\times 10^{-9}$ & $10^{-8}$ & $7.5\times10^{-6}$ & $5.2\times 10^{-9}$ & $8.1\times 10^{-9}$ & $1.8\times10^{-5}$ & - & - & - \\
 & 1 & $8.2\times10^{-11}$ & $5.6\times10^{-10}$ & $3.4\times10^{-10}$ & $4.1\times10^{-10}$ & $2.1\times10^{-9}$ & $6.1\times10^{-8}$ & $1.5\times10^{-9}$ & $5.7\times 10^{-9}$ & $1.1\times10^{-6}$ & $4.3\times 10^{-9}$ & $1.1\times 10^{-8}$ & $7.1\times10^{-6}$ & $1.2\times 10^{-8}$ & $2.3\times 10^{-8}$ & $3.4\times10^{-5}$ & $1.2\times 10^{-8}$ & $1.3\times 10^{-8}$ & $5.5\times10^{-5}$ & - & - & - \\
\hline

 & 0.05 & $5.0\times10^{-14}$ & $10^{-12}$ & 0.0 & $6.0\times10^{-13}$ & $9.0\times10^{-12}$ & 0.0 & $4.6\times10^{-12}$ & $3.6\times10^{-11}$ & $1.1\times10^{-11}$ & $1.9\times10^{-11}$ & $9.1\times10^{-11}$ & $5.6\times10^{-9}$ & $7.9\times10^{-11}$ & $2.3\times10^{-10}$ & $6.3\times10^{-8}$ & - & - & - & - & - & - \\
2\,h & 0.2 & $1.5\times10^{-12}$ & $1.5\times10^{-11}$ & 0.0 & $9.6\times10^{-12}$ & $6.0\times10^{-11}$ & $6.1\times10^{-11}$ & $3.8\times10^{-11}$ & $1.6\times10^{-10}$ & $6.2\times10^{-9}$ & $1.2\times10^{-10}$ & $3.2\times10^{-10}$ & $6.0\times10^{-8}$ & $3.9\times10^{-10}$ & $6.5\times10^{-10}$ & $3.1\times10^{-7}$ & - & - & - & - & - & - \\
 & 1 & $1.7\times10^{-11}$ & $7.3\times10^{-11}$ & 0.0 & $6.6\times10^{-11}$ & $2.1\times10^{-10}$ & $6.8\times10^{-11}$ & $1.9\times10^{-10}$ & $4.5\times10^{-10}$ & $6.4\times10^{-9}$ & $4.0\times10^{-10}$ & $6.8\times10^{-10}$ & $6.1\times10^{-8}$ & $7.6\times10^{-10}$ & $9.6\times10^{-10}$ & $3.1\times10^{-7}$ & - & - & - & - & - & - \\
\hline

\end{tabular}
}
\caption{Average number of active QPEs $\langle N_\mathrm{QPE} \rangle$ over $10$ Gyr, as estimated from two-body relaxation simulations. We use the same symbols as in Table \ref{table:1}. At each snapshot of the simulation, we compute $N_\mathrm{QPE}=N_\mathrm{disks} \, N_\mathrm{CO}$ as explained in Sec.~\ref{sec:qpe_abundance}. ``-'' means that the period at $r_\mathrm{cap}$ is longer than $T_\mathrm{QPE}$.}
\label{table:2}
\end{table}

\end{appendix}
\end{landscape}

\end{document}